\documentclass[10pt,preprint]{aastex}

\usepackage{graphics}
\usepackage{epsfig}
\usepackage{amssymb}
\usepackage{rotating}
\usepackage{lscape}
\usepackage{amsmath}

\shorttitle{Angular momentum loss from cool stars}
\shortauthors{Barnes \& Kim}
\slugcomment{Journal reference: Barnes \& Kim, 2010, ApJ, 721, 675}

\begin{document}

\title{Angular momentum loss from cool stars: 
\\an empirical expression and connection to stellar activity}

\author{Sydney A. Barnes}
\affil{Lowell Observatory, Flagstaff, AZ, USA}

\author{Yong-Cheol Kim\altaffilmark{1}}
\affil{Astronomy Department, Yonsei University, Seoul, South Korea}

\altaffiltext{1}{Visiting Astronomer, Lowell Observatory, Flagstaff, AZ, USA}


\begin{abstract} \label{abs}

We show here that the rotation period data in open clusters allow the
empirical determination of an expression for 
the rate of loss of angular momentum from cool stars on the main sequence. 
One significant component of the expression, the dependence on rotation rate,
persists from prior work; others do not.
The expression has a bifurcation, as before, that corresponds to an observed 
bifurcation in the rotation periods of coeval open cluster stars.
The dual dependencies of this loss rate on stellar mass are captured by two 
functions, $f(B-V)$ and $T(B-V)$, that can be determined from the rotation 
period observations. 
Equivalent masses and other [$UBVRIJHK$] colors are provided in Table\,1.
Dimensional considerations, and a comparison with appropriate calculated 
quantities suggest interpretations for $f$ and $T$, both of which appear to 
be related closely (but differently) to the calculated convective turnover 
timescale, $\tau_c$, in cool stars.
This identification enables us to write down symmetrical expressions
for the angular momentum loss rate and the deceleration of cool stars,
and also to revive the convective turnover timescale as a vital connection 
between stellar rotation and stellar activity physics.
\end{abstract}



\section{Introduction} \label{intro}

Cool stars are known to lose angular momentum and spin down over time.
However, a detailed understanding of this loss has yet to be achieved.
Measurements from the ground and space, coupled with theory, give us
a good idea of the rate of angular momentum loss for the present-day Sun, 
but extensions to stars of different masses or those of other ages, 
particularly very young stars, are problematical in various ways. 
This paper proposes a path toward making such extensions. 

In the same paper that proposed the existence of the solar wind, Parker (1958) 
noted that this wind would cause a `retardation of solar rotation' on a 
timescale similar to its lifetime. 
Weber \& Davis (1967) elaborated on this particular effect and showed 
that the associated angular momentum loss rate is equivalent to assuming 
corotation of the solar wind out from the surface to the Alfvenic radius, $r_A$.
Schatzman (1962) had previously suggested that the solar case could be 
generalized 
to other stars, linked the associated angular momentum loss to the presence of 
a surface convection zone, and provided a formula for the loss rate in terms of 
a star's angular velocity and certain other quantities. However, it was not 
clear how to calculate or measure these quantities. A similar criticism could 
be leveled against the expression for angular momentum loss provided by 
Mestel (1968). The associated viewpoint that the angular momentum loss rate,
${dJ/dt} \propto \Omega r_A^2 \dot{M}$ where $\Omega$ and $\dot{M}$ are the
star's angular velocity and mass loss rate respectively, ultimately runs into
the difficulty of providing credible calculations or measurements of both
$r_A$ and $\dot{M}$ for the Sun and cool stars of other masses. Alternatively,
writing $dJ/dt \propto B_0^2 R^4 \Omega/v_A$, where $B_0$, $R$, and $v_A$ are
the surface magnetic field, stellar radius, and wind velocity at the Alfven
surface, requires deriving or measuring $B_0$ and $v_A$ as a function of a
star's mass and other properties. See Collier Cameron \& Li (1994) for a 
discussion of the issues that arise when one takes this viewpoint.

Another way to approach this problem is empirical. 
Kraft (1967) measured a decline of stellar rotation velocities with age, 
strengthening the solar-stellar analogy, and Schatzman's work provided a 
natural framework for interpreting the break in the Kraft (1967) curve of 
rotation velocities against stellar mass. 
The work of Skumanich (1972) provided more direct information about the angular 
momentum loss rates for solar-mass stars when he noted that their measured 
rotation velocities, $v \sin i$, decline with age, $t$, as 
\begin{equation}
\overline{ v \sin i} \propto t^{-0.5}
\end{equation}
where $\overline{v \sin i} $ represents an average over coeval cluster stars.
This observation implies that the rate of loss of angular momentum for 
(solar-mass) stars of constant structure (i.e. moment of inertia) obeys the 
relationship
\begin{equation}
\frac{dJ}{dt} \propto - \Omega^3.
\end{equation}
According to this viewpoint, if the observations can provide the acceleration, 
$d\Omega/dt$ as a function of all relevant variables, then $dJ/dt$ immediately 
follows if the associated moment of inertia can be inferred.

A hybrid approach helps overcome certain weaknesses of both viewpoints.
Endal \& Sofia (1981) used the wind-loss formulation of 
Belcher \& MacGregor (1976), in turn based on the work of Weber \& Davis (1967), 
to understand the rotational history of the Sun, 
but the resulting rotational evolution (their Figure\,4), including certain
assumptions about internal transport of angular momentum, did not closely 
match the empirical results of Skumanich (1972). At that time, the security of 
the Skumanich result was not assured, and its applicability was restricted
to stars of near-solar mass. Indeed, Mestel (1984) investigated other 
loss relationships than $dJ/dt \propto - \Omega^3$ and made significant steps 
toward elucidating what they implied.
Kawaler (1988) extended Mestel's work in certain significant ways, identifying 
dependencies in addition to $\Omega$ in the expression for angular momentum 
loss, and proposed an expression that could be implemented easily in models of 
rotating cool stars of {\em varying} mass. 
The relationship, combining theoretical and observational considerations, is
\begin{equation}
\frac{dJ}{dt} = - K_w \Omega^{1+4an/3} (\frac{R}{R_{\odot}})^{2-n} 
                      ({\dot M_{14}})^{1-2n/3} (\frac{M}{M_{\odot}})^{-n/3}
\end{equation}
where $R$, $M$, and ${\dot M}$ are the stellar radius, mass, and mass loss
rate respectively, 
$n$ parameterizes the geometry of the magnetic field, 
$a$ is the exponent that relates the surface magnetic field,
$B_0$ to $\Omega$ via $B_0 \propto \Omega^a$, 
and $K_w$ is a calibration constant chosen to ensure that a solar model attains
the solar rotation rate at solar age. 
$n$ can vary from $3/7$ for a dipolar field to $2$ for a purely radial field.
If one assumes a linear dynamo, then $a = 1$, 
and $n$ must then be set equal to $1.5$ to reproduce the observed Skumanich 
spin-down, $\overline{v \sin i} \propto t^{-0.5}$. 
This choice of $n$ kills the mass loss term, of course.
Consequently, the above expression usually simplifies to
\begin{equation}
\frac{dJ}{dt} = - K_w \Omega^{3} (\frac{R}{R_{\odot}})^{0.5} 
                      (\frac{M}{M_{\odot}})^{-0.5},
\end{equation}
eliminating the mass loss rate, ${\dot M}$.
This expression, a hybrid of the two viewpoints discussed above, has been 
routinely used to drain angular momentum from rotational stellar models 
constructed using YREC, the Yale Rotating Stellar Evolution Code 
(e.g., Pinsonneault et al. 1989), 
and also is used in other stellar models (e.g., Bouvier et al. 1997).
A didactic account of the above developments, set in a broader context than
here, may be found in Chapters 13 and 21 of Maeder (2009).

We note that some studies of the rotational evolution of stars do not 
explicitly provide an expression for angular momentum loss
(e.g., MacGregor \& Brenner 1991; Armitage \& Clarke 1996). 
This is usual in cases where the magnetic field
and associated quantities (and sometimes their evolution) are themselves 
modeled numerically. The mass loss rate is usually an input parameter.
A recent example is the work of Matt \& Pudritz (2008), 
which provides details of such an approach. 
Some models include the possible effects of disks on the pre-main-sequence
(e.g., Collier Cameron \& Campbell 1993, Keppens et al. 1995, 
Sills et al. 2000, Barnes et al. 2001). 
But here we are concerned only with the main-sequence where any possible 
effect of pre-main-sequence disks has abated. 
We also note that cataclysmic variable (CV) research uses an expression 
derived by Rappaport et al. (1983) for the extra-gravitational loss of angular 
momentum. This expression has some similarity to the Kawaler (1988) 
expression, as noted by Andronov et al. (2003). 
However, CVs are beyond the scope of this paper.

In the observational domain, rotational data for stars in open clusters 
had revealed (and further observations confirmed) the presence of the so-called 
ultra-fast rotators (UFRs) in the Pleiades (van Leeuwen \& Alphenaar 1982).
These observations implied that the angular momentum loss rate for such stars 
was {\em much lower} than that calculated using Kawaler's expression.
Chaboyer et al. (1995), following MacGregor \& Brenner (1991), 
therefore suggested modifying this 
relationship at high rotation rates to facilitate the theoretical modeling 
of such stars. This idea became known as ``saturation'' 
(e.g., Stauffer 1994)
and was roughly parallel to a similar phenomenon in soft X-rays, 
(although the exact connection between the two was not clear then).
Chaboyer et al. (1995) suggested the loss relationship
\begin{equation}
\frac{dJ}{dt} = - K_w (\frac{R}{R_{\odot}})^{0.5} (\frac{M}{M_{\odot}})^{-0.5} 
       \begin{cases} 
            \hspace{0.8cm} \Omega^3  & \text{for $\Omega \leq \Omega_{\rm crit}$,} \\
            \Omega_{\rm crit}^2 \, \Omega  & \text{for $\Omega > \Omega_{\rm crit}$,}
       \end{cases}
\end{equation}
where $\Omega_{\rm crit}$ is a constant.
(A way of achieving a similar effect in the context of Weber-Davis type
wind models is presented in Collier Cameron \& Li 1994.)
Furthermore, Barnes \& Sofia (1996) showed that the UFRs could not be modeled
by the original Kawaler expression, with its $\Omega^3$ dependence, 
regardless of any possible pre-main-sequence spin-up, 
the bloated state of T\,Tauri stars being considered a possible
reservoir of angular momentum suited to explain the origin of the UFRs.
As a result, subsequent work, e.g., Krishnamurthi et al. (1997) and 
Barnes et al. (2001) 
routinely used the newer Chaboyer et al. (1995) loss prescription. 
It was hoped that $\Omega_{\rm crit}$ could simply be set equal to some 
constant threshold angular velocity for all relevant cool stars.

However, the accumulation of additional data suggested that a single constant 
value of $\Omega_{\rm crit}$ for all cool stars was inadequate
(Barnes \& Sofia 1996; Krishnamurthi et al. 1997).
Consequently, Krishnamurthi et al. (1997) advocated a scaling 
\begin{equation}
\Omega_{\rm crit} = \Omega_{\rm crit \odot} \frac{\tau_{\odot}}{\tau_c} 
\end{equation}
where $\tau_c$ is the convective turnover timescale. 
(Prior to this, and in the context of using a Weber-Davis type wind to 
understand the rotational evolution of stars, Collier Cameron \& Li (1994) 
had also suggested a scaling involving the convective turnover timescale, 
in this case setting the 
star's surface magnetic field, $B_0$ according to $B_0 \propto \tau_c \Omega$.)

The Ohio State University group 
(Sills et al. 2000; Andronov et al. 2003; Denissenkov et al. 2010)
absorb the solar radius and mass from the Kawaler/Chaboyer formula into the 
wind constant, $K_w$, and write:
\begin{equation}      \label{eqn:Jloss_SPT}
\frac{dJ}{dt} = - K_w \sqrt{\frac{R}{M}} 
         \begin{cases}
             \hspace{0.8cm} \Omega^3  & \text{for $\Omega \leq \Omega_{\rm crit}$,} \\
             \Omega_{\rm crit}^2 \, \Omega   & \text{for $\Omega > \Omega_{\rm crit}$,}
         \end{cases} 
\end{equation}
but $\Omega_{\rm crit}$ is now variable as described immediately above, or
varied piece-wise to match open cluster observations. For example,
Sills et al. 2000 used the Krishnamurthi et al. (1997) scaling for 
$\Omega_{\rm crit}$ 
for masses down to $0.6 M_{\odot}$, and set $\Omega_{\rm crit}$ 
manually for each modeled stellar mass below this. 
Stars of very low mass are beyond the scope of this work, but we note that such
a tuning effectively modifies $\Omega_{\rm crit}$ from a constant into a new 
and arbitrarily modifiable function.
Making $\Omega_{\rm crit}$ mass dependent implies, of course, that the 
$\sqrt{R/M}$ term in the expression for $dJ/dt$ is simply not capturing the 
mass dependence of the angular momentum loss, as it was originally intended to. 
Clearly, there is a problem.

On the other hand, the bifurcation in the angular momentum loss, as proposed by MacGregor \& Brenner (1991), encapsulated in the relationship of 
Chaboyer et al. (1995),
 and as subsequently used in modeling is important in retrospect, because 
it does capture an essential feature of the observations.
Indeed, since then, a steadily growing rotation period database has allowed the 
identification of distinct fast (C-) and slow (I) sequences in color-period
diagrams of open cluster stars, as initially proposed by Barnes (2003).
This C/I classification has been confirmed by extensive rotation period
observations in 
M\,35 (Meibom et al. 2009), M\,37 (Hartman et al. 2009) and 
M\,50 (Irwin et al. 2009), 
the first
including a decade-long radial velocity survey for cluster membership and
multiplicity. Scholz \& Eisloffel (2007) include a discussion of this bifurcation
vis-a-vis rotation periods in Praesepe/Hyades, and most recently, data in
Hartman et al. (2010) clearly display this bifurcation in a large and uniform 
rotation period study of the Pleiades. 

This observed bifurcation ties in well with the bifurcation in the angular 
momentum loss expression proposed by Chaboyer et al. (1995), although not to 
the mass dependence used there, which is the same for both kinds of stars.
We shall see below that the mass dependence is actually different for 
the two sequences, so that it is {\it impossible} for the same expression to
describe both.

Barnes (2003) also suggested interpretations of the observed shapes
of the C- and I\,sequences, based on theoretical considerations, and a unifying 
scenario (hereafter called the CgI scenario) proposed in that work for the 
rotational evolution of cool stars. In particular, he suggested that stars 
initially are fast rotators on the C\,sequence, where the inner radiative and 
outer convection zones are largely decoupled, so that the mass dependence of
this sequence is specified by (the reciprocal of) the 
moment of inertia of the outer convection zone.
The observed transition of stars from the C- to the I\,sequence was proposed
to be coincident with a change in the mass dependence, which was itself proposed
to change from that of the outer convection zone alone to that of the entire
star, and also to be coincident with the onset of an interface dynamo.
Thus the mass dependence for the I\,sequence was suggested to be dependent on
(the reciprocal of the square root of) the moment of inertia of the whole star.

This work begins by considering and checking whether these proposed 
dependencies work. We show that they do not (Section 2).
We then show that the observations themselves might be queried to provide 
$dJ/dt$ directly in terms of observed quantities (Section 3).
Section 4 proposes an interpretation of the relevant observed quantities in 
terms of the convective turnover timescale. 
The relationship with stellar activity physics is pointed out in Section 5, 
and the conclusions are stated immediately thereafter.
(The next paper in this series combines the C- and I-type behaviors identified 
in Section 4 into a simple nonlinear model for the rotational evolution of 
stars and explores the consequences for gyrochronology.)

\section{Inadequacy of the moment of inertia proposal}

We begin by negating a proposal made by Barnes (2003),
that the mass dependence of the rotation periods in open clusters 
can be simply attributed to the moments of inertia of either
the whole star or that of the surface convection zone.

    \subsection{I\,sequence}

Kawaler (1989)
was the first to note that beyond an age of a few hundred million 
years, as exemplified by the 600\,Myr-old Hyades open cluster, a deterministic 
relationship between rotation period, color and age could be derived, and 
inverted to provide a star's age. Younger clusters have a more complicated
morphology, initially parsed into fast/C- and slow/I sequence stars by
Barnes (2003).
He proposed that the latter, I\,sequence stars are describable
by $P_I = f(B-V) \times g(t)$ where $f(B-V)$ and $g(t)$ are empirically 
determinable functions of the $B-V$ color and age, $t$, respectively. 
In that paper, the functional forms used were 
$f(B-V) = \sqrt{(B-V-0.5)}-0.15(B-V-0.5)$ and $g(t)=\sqrt{t}$.
These functions have been subsequently re-determined, most notably by
Meibom et al. (2009), based on a very large study of both rotation periods
and membership in the open cluster M\,35. They determined that
$f(B-V) = 0.77 (B-V-0.47)^{0.55}$, 
and for definiteness, we will use this latter form in this paper\footnote{In a later comparison, we will also show the Mamajek \& Hillenbrand (2008) determination of $f(B-V)$. We also acknowledge that the index $n$ in $g(t) = t^n$ is usually found to exceed $0.5$ slightly (e.g., Barnes 2007, Collier Cameron et al. 2009, James et al. 2010).}. 
(Transformation to mass or other colors in the set [$UBVRIJHK$] can be 
accomplished using Table\,1.)

Barnes (2003) suggested identifying $f(B-V)$ with $1/\sqrt{I_*}$, 
where $I_*$ is the moment of inertia of the star. We have calculated $I_*$ 
using the latest version of YREC, and display it in Figure\,1 as a 
function of mass and of $B-V$ color for a series of $500$\,Myr models of 
solar composition. This age was selected to ensure that all 
lower-than-solar-mass stellar models of interest have passed through the 
pre-main-sequence phase and arrived on the main-sequence, while 
higher-than-solar-mass models have not evolved off the main-sequence. 
To the precision of this work, further main-sequence evolution does not have an
appreciable effect. The numerical values and associated [$UBVRIJHK$] colors 
using both Green et al. (1987) and Lejeune et al. (1997), Lejeune et al. (1998) color transformations are provided in Table\,1. 

\begin{figure}[h]     
\begin{minipage}[b]{0.5\linewidth} 
\centering
\includegraphics[width=5.75cm,angle=-90]{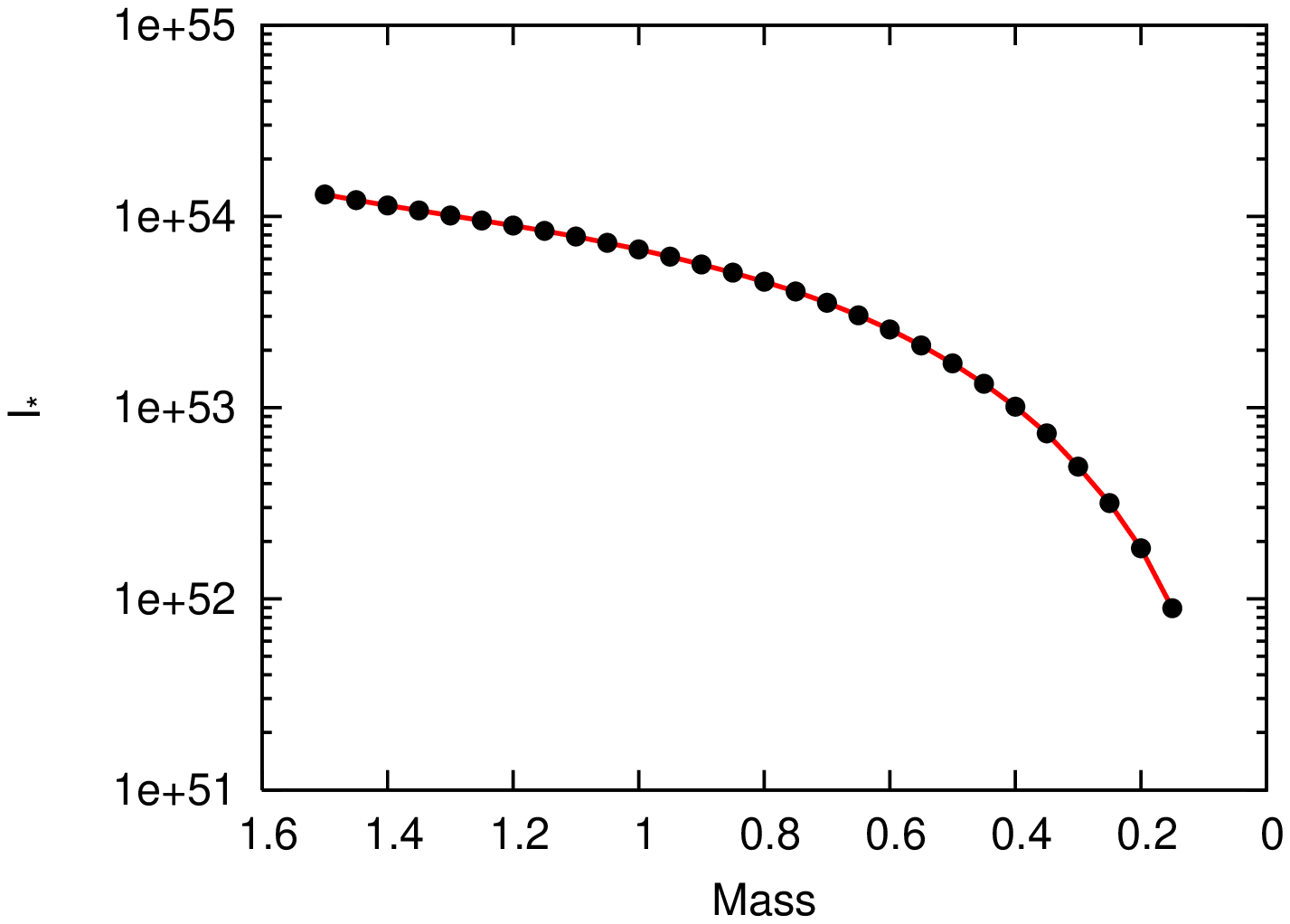}
\end{minipage}
\begin{minipage}[b]{0.5\linewidth}
\centering
\includegraphics[width=5.75cm,angle=-90]{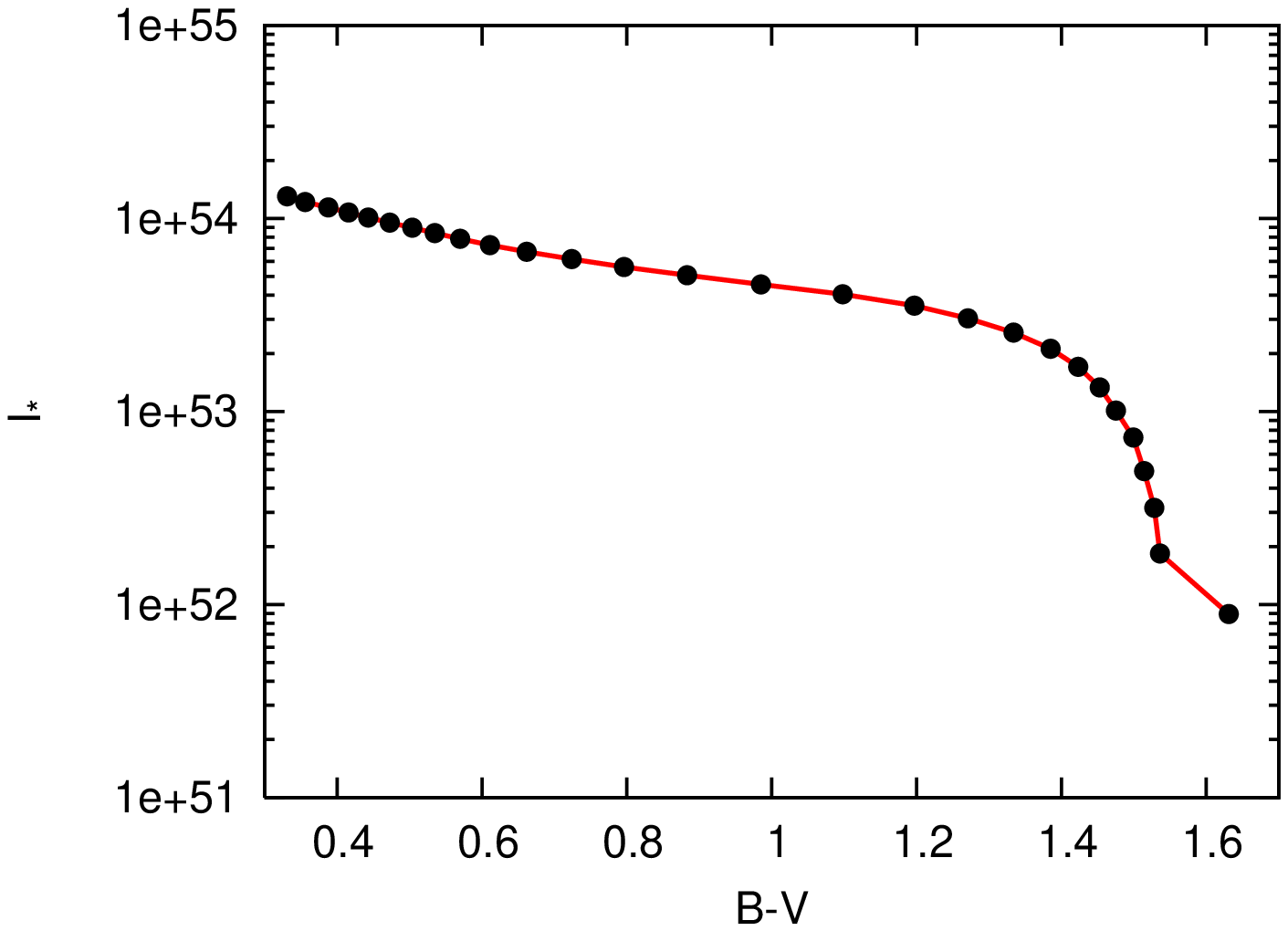}
\end{minipage}
\caption{ \small 
{\bf Left:} The moment of inertia of the entire star, $I_*$, against stellar 
mass, in the range $1.5-0.15 M_{\odot}$. (Note the inverted mass axis.)
{\bf Right:} The same against $B-V$ color. 
As expected, $I_*$ increases monotonically with stellar mass.
(In this, and subsequent figures, we display the Lejeune et al. (1997, 1998)
 $B-V$ color; other colors in the set [$UBVRIJHK$], including those with the
 Green et al. (1987) color transformations are provided in Table\,1.)
}
\end{figure}

We compare $f(B-V)$ with $1/\sqrt{I_*}$ in Figure\,2.
The curves are normalized to agree for the solar case.
It is clear that the agreement is not good.
One could argue that the normalization makes it hard to tell how poor the fit 
is for the cooler stars. The key part of the disagreement, however, is
that $1/\sqrt{I_*}$ does not drop off sufficiently fast for the warmer stars.
This disagreement would only get worse if the normalization were changed.

\begin{figure}[h]	   
\includegraphics[scale=0.60,angle=-90]{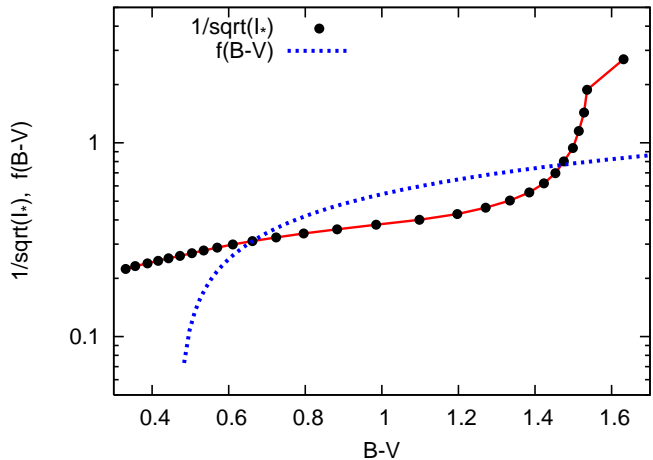}
\caption{ \small 
Comparison of the empirical function $f(B-V)$ for the I\,sequence (dashed blue),
with $1/\sqrt{I_*}$ (filled circles). The latter was proposed by Barnes (2003) 
to interpret the empirical dependence. 
(The $I_*$ curve is normalized for solar mass.)
We conclude that the two curves are not as similar as was proposed in 
Barnes (2003), negating the corresponding identification. 
In particular, $1/\sqrt{I_*}$ does not drop steeply enough 
to model the observed rotation period decline among the warmer stars.
}
\end{figure}

    \subsection{C\,sequence}

Barnes (2003) proposed that the C\,sequence of faster rotating stars
can be described by $P_C = P_0 e^{t/T(B-V)}$, where $P_C$ denotes the rotation
periods of the stars in question, $t$ is the age, and $T(B-V)$ is the
relevant spin-down timescale\footnote{The choice of an exponential function in $t$ is partly motivated by theoretical considerations, and it is possible that other functions of $t$ are also suitable, as discussed below. $T(B-V)$ is our main concern here.}.
The open cluster observations indicate that $T(B-V)$ is short for stars
bluer than the Sun, tending toward zero at mid-F.
On the other side, redward of the Sun, $T(B-V)$ is known to increase.
The main requirement is a function that dives to zero at $B-V = 0.47$,
the x-intercept of $f(B-V)$, as determined by Meibom et al. (2009).
The most reliable empirical determinations of $T(B-V)$ for G and K stars are 
also by Meibom et al. (2009), who find that 
$T(B-V)$ = 60\,Myr for G stars (specifically, those with $ 0.6 < B-V < 0.8 $) 
and 
$T(B-V)$ = 140\,Myr for K stars, (with $ 0.8 < B-V < 1.3 $).
We estimate uncertainties on these values of 5 Myr and 13 Myr respectively, and
use these values to represent those for  mean $B-V$ colors of 0.7 and 1.05.
For redder stars, we note that about a third of the Hartman et al. (2009) 
sample of stars from the 550\,Myr-old open cluster M37 are on the C\,sequence.
This implies that $T(B-V)$ = 500$\pm$65\,Myr for early M stars, represented by 
a point at $B-V$ = 1.45.
$T(B-V)$ is plotted using these discrete values in Figure\,3. We also 
display a cubic spline fit to these data, and acknowledge that $T(B-V)$
becomes increasingly uncertain with decreasing stellar mass\footnote{The uncertainties quoted above are based on fractional numbers of stars, and do not include other effects such as uncertainties in the assumed cluster isochrone ages.}.
Similar results - $50-100$\,Myr for $1 M_{\odot}$, $0.5-1$\,Gyr for 
$0.5 M_{\odot}$, and several Gyr for $M < 0.3 M_{\odot}$ -
were quoted by Scholz et al. (2009) and references therein.

\begin{figure}[h]	   
\includegraphics[scale=0.60,angle=-90]{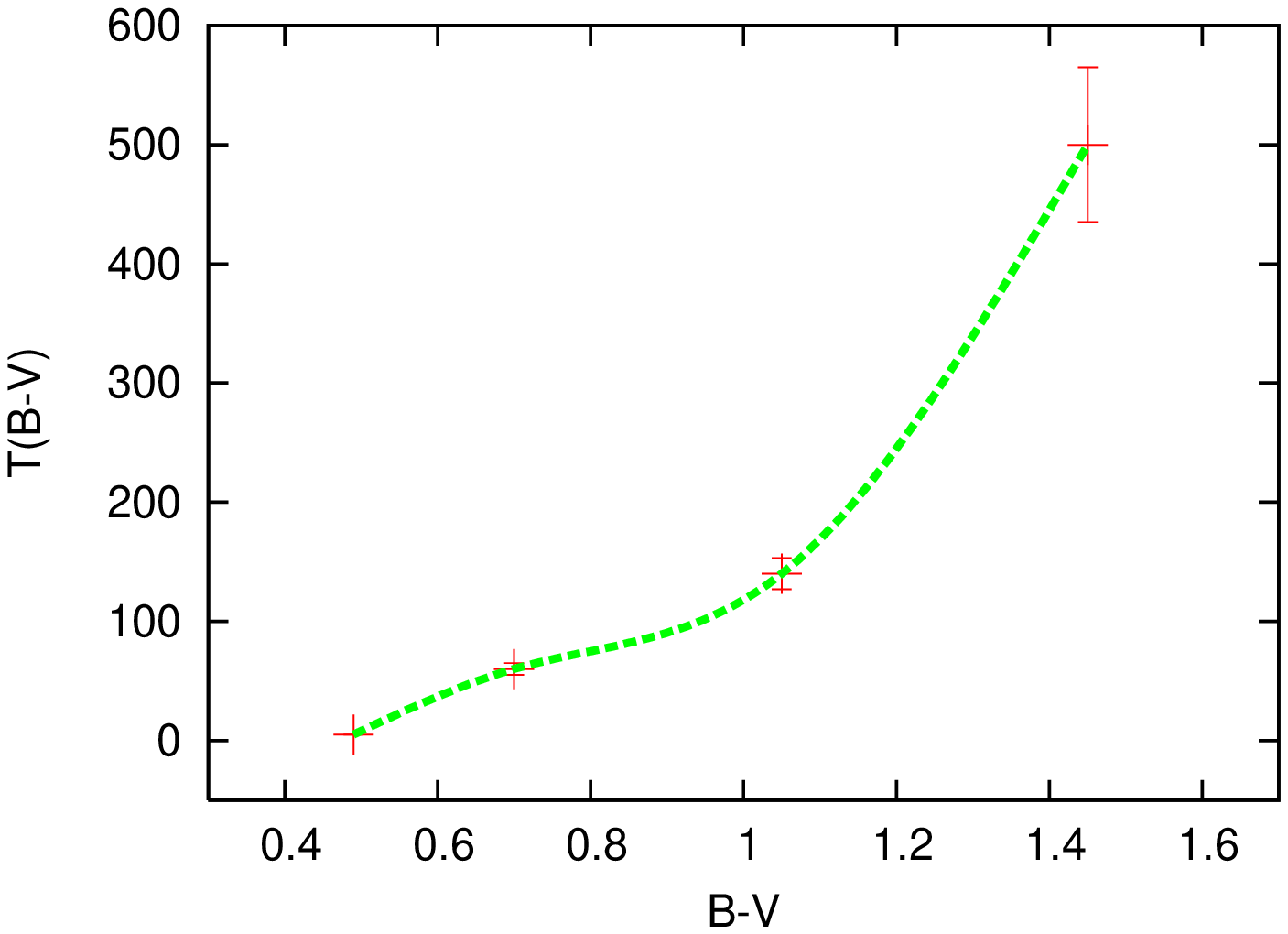}
\caption{ \small 
The C\,sequence spin-down timescale, $T(B-V)$, for cool stars (in units of Myr) 
plotted against $B-V$ color. 
The green line is a cubic spline fit to the data.
$T(B-V)$ increases steadily with $B-V$ color, somewhat faster than linear
for the coolest stars. The data points are discussed in the text.
}
\end{figure}

Barnes (2003) suggested that $T(B-V) \propto I_{cz}$, 
where $I_{cz}$ is the moment of inertia of the convection zone.
We have calculated $I_{cz}$ using YREC, and display it in Figure\,4 as a 
function of mass and of $B-V$ color for a series of $500$\,Myr models of solar
composition. $I_{cz}$ drops at both ends as expected; it decreases for stars
with masses lower than $0.8\,M_{\odot}$, and also for the warmer stars with
thinning surface convection zones.
The associated numerical values and other colors are provided in Table\,1.

\begin{figure}[h]     
\begin{minipage}[b]{0.5\linewidth} 
\centering
\includegraphics[width=5.75cm,angle=-90]{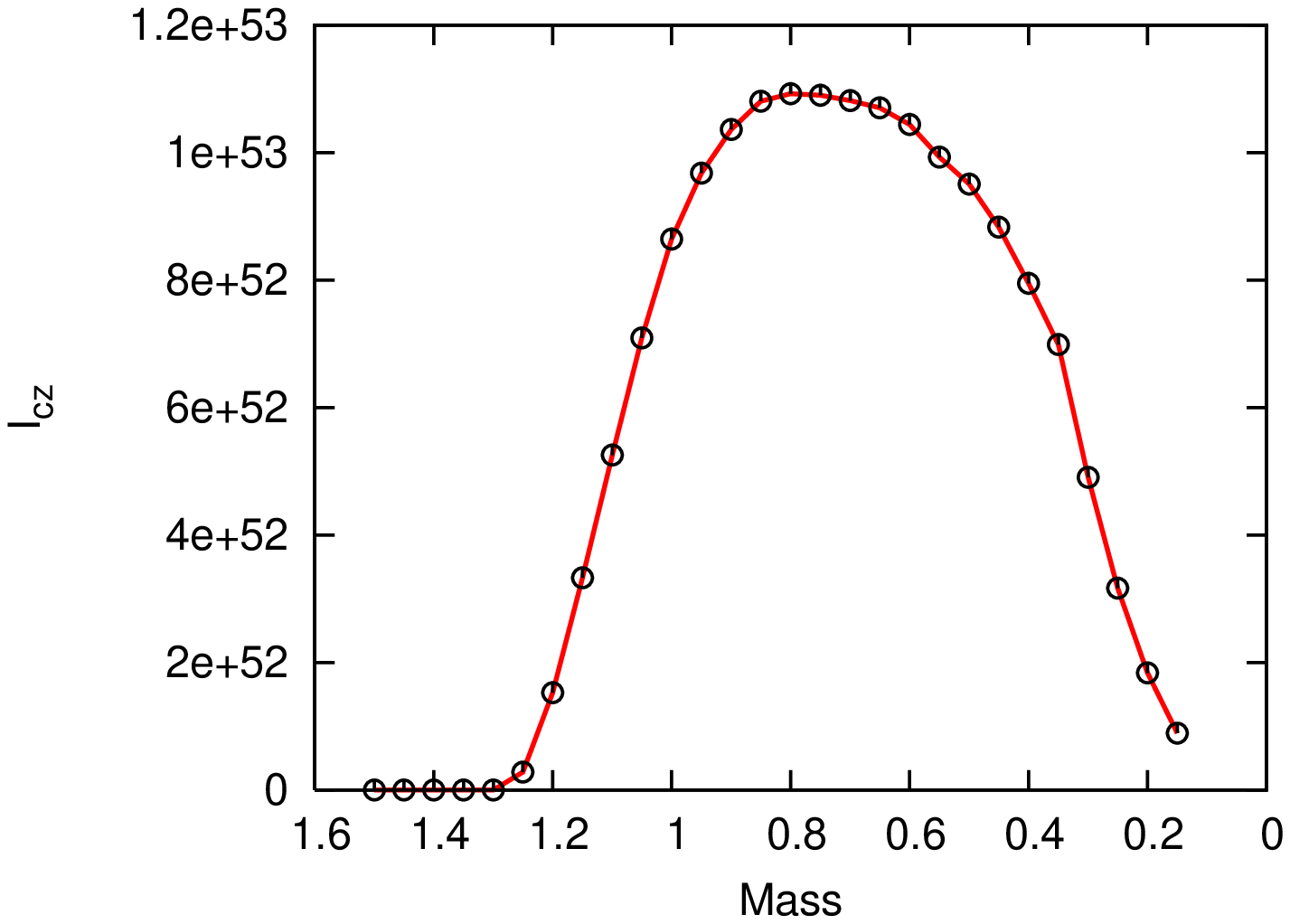}
\end{minipage}
\begin{minipage}[b]{0.5\linewidth}
\centering
\includegraphics[width=5.75cm,angle=-90]{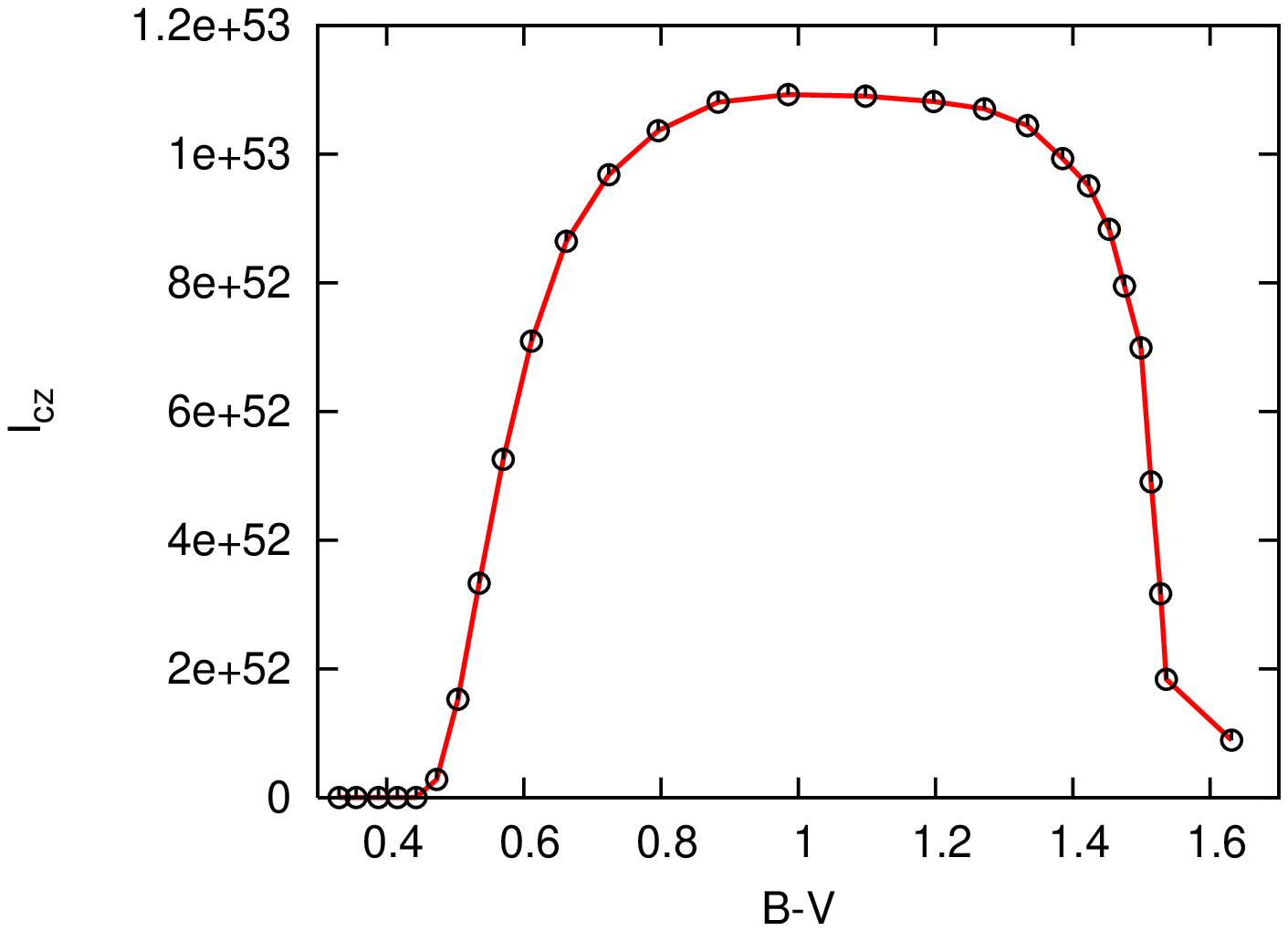}
\end{minipage}
\caption{ \small 
{\bf Left:} The moment of inertia of the outer convection zone, $I_{\rm cz}$, 
against stellar mass. (Note the inverted mass axis.)
{\bf Right:} The same against $B-V$ color. 
$I_{\rm cz}$ drops at both ends as expected; to the left as surface convection 
zones thin out and vanish, and to the right, with diminishing stellar mass.
}
\end{figure}

We compare $T(B-V)$ with $I_{cz}$ in Figure\,5.
The two variables are in reasonable agreement for $M > M_{\odot}$ but diverge 
badly for 
$M < M_{\odot}$. Whereas $T(B-V)$ is a steadily increasing function of $B-V$ 
(or decreasing stellar mass), $I_{cz}$ becomes relatively flat for K dwarfs,
and indeed drops for M dwarfs, as might be expected. 
Consequently, we find that $I_{cz}$ is a poor match to $T(B-V)$,
and withdraw the corresponding suggestion made in Barnes (2003).

\begin{figure}[h]	   
\includegraphics[scale=0.60,angle=-90]{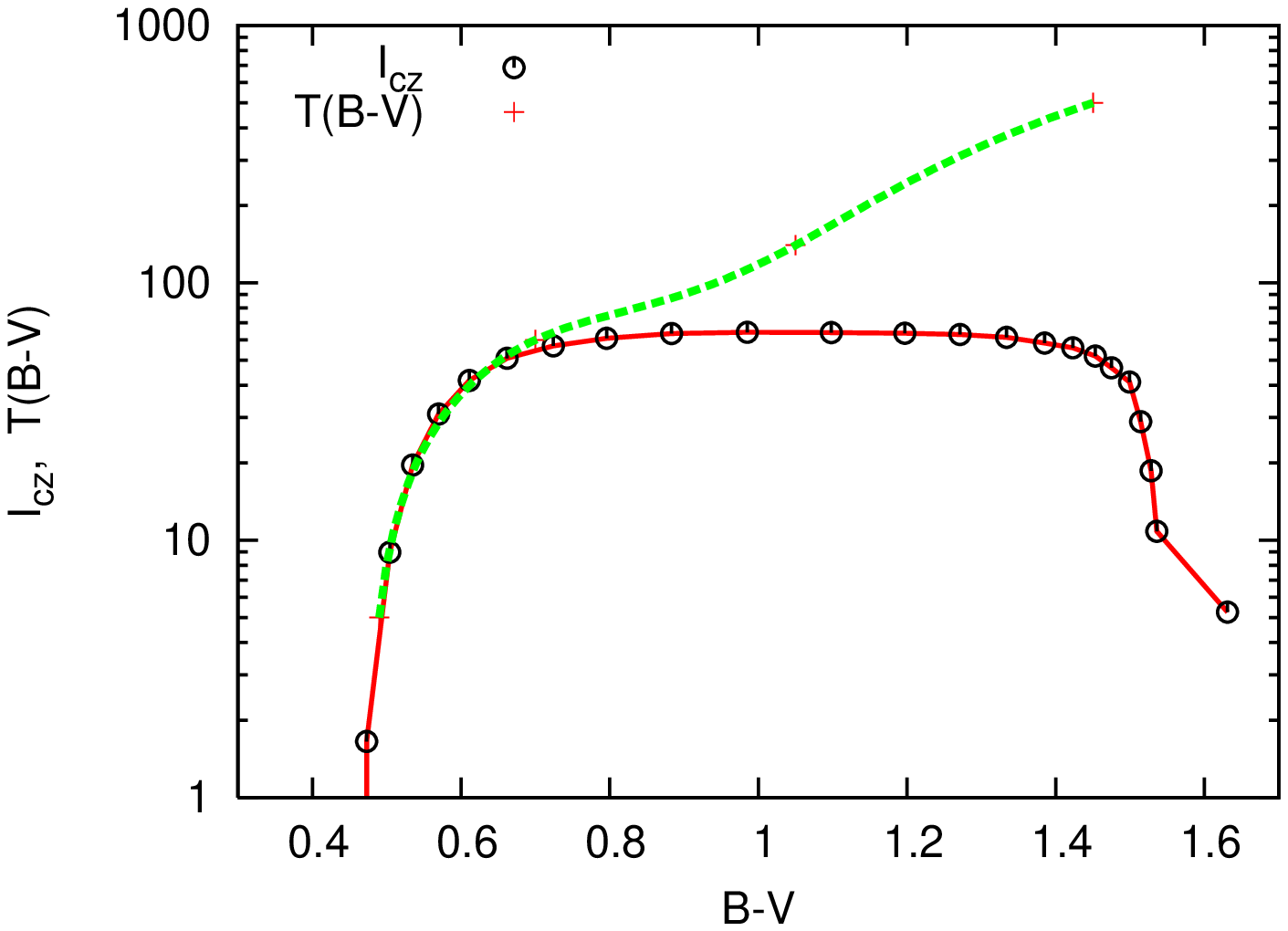}
\caption{ \small 
The C\,sequence spin-down timescale $T(B-V)$, in units of Myr (pluses; 
dashed green line), is compared with $I_{\rm cz}$, the moment of inertia of the 
surface convection zone in cool stars (unfilled circles; solid red line). 
$I_{\rm cz}$ is normalized for solar mass.
Although the agreement is good for $M > M_{\odot}$, partly because both 
$I_{\rm cz}$ and $T$ drop to zero at similar color, $I_{\rm cz}$ and $T(B-V)$ can 
be observed to diverge for lower mass stars, negating the corresponding 
identification proposed in Barnes (2003).
}
\end{figure}

\section{Empirical approach}

It transpires that the observations themselves can be queried to provide an
empirical loss rate, through a logical extension of prior work. This is 
accomplished here, considering first the I\,type rotators, and then the C\,type 
ones.

\subsection{I\,sequence}

We return to the observational basis that the rotation periods, $P_I$, 
of stars on the I\,sequence are describable as a product of two separable 
functions, $f$, and $g$, of the star's mass or $B-V$ color and age, $t$, 
respectively:
\begin{equation} \label{eqn:P_I}
P_I = f(B-V) \times g(t).
\end{equation}
This was shown by construction in Barnes (2003) and Barnes (2007), 
where both $f(B-V)$ and $g(t)$ were determined empirically. 
\begin{equation} \label{eqn:g(t)}
g(t) = \sqrt{t}
\end{equation}
was found there to be applicable to a wide range of cool stars,
a result tracing back to the work of Skumanich (1972),
where the more restrictive case of solar-mass stars was considered. 
As in the prior section, we will use the Meibom et al. (2006) determination 
\begin{equation} \label{eqn:f(B-V)} 
f(B-V) = 0.77 (B-V-0.47)^{0.55} 
\end{equation}
over other determinations, including those of Barnes (2003) and 
Barnes (2007),
because of the size and quality of the data set considered there.

Changing variable from $P_I$ to $\Omega$ via $P_I = 2 \pi/ \Omega$,
equation (8) becomes
\begin{equation} \label{eqn:Omega}
\frac{2 \pi} {\Omega} = f(B-V) \times g(t), 
\end{equation}
and upon differentiating it with respect to time, we get
\begin{equation}
\frac{d \Omega}{dt} = - \frac{\Omega^2}{2\pi} f(B-V) \frac{dg}{dt}. 
\end{equation}
Using $g(t) = \sqrt{t}$ from Equation (9), we get
\begin{equation}
\frac{d \Omega}{dt} = - \frac{\Omega^2}{4\pi} f(B-V) \frac{1}{\sqrt{t}}.
\end{equation}
Re-using Equation (11) to write $\sqrt{t}$ in terms of 
$f$ and $\Omega$, we get
\begin{equation} \label{eqn:dO_Idt}
\frac{d\Omega}{dt} = - \frac{\Omega^3}{8 {\pi}^2} f^2(B-V),
\end{equation}
which specifies the deceleration of the star in purely observational
quantities. Note that (the square of) $f(B-V)$ completely specifies the mass 
dependence. $f(B-V)$ being zero for mid-F and 
earlier-type stars, the deceleration is correspondingly zero for these stars. 
$f(B-V)$ increases steadily through late-F, G, K, and early\,M stars, so that 
the deceleration of these stars is correspondingly greater. 
This accounts for the shape of the I\,sequence in open clusters.
It is also worth emphasizing that the open cluster data directly provide only 
the deceleration above, not the angular momentum loss rate, $dJ/dt$, 
derived below. 

Deriving the rate of angular momentum loss requires an additional step.
By the Chain Rule,
\begin{equation}
 \frac{dJ}{dt} = I \frac{d\Omega}{dt} + \Omega \frac{dI}{dt}
\end{equation}
so that simple substitution from Equation (14) above gives
\begin{equation}
 \frac{dJ}{dt} = - \frac{\Omega^3}{8 \pi^2} I f^2(B-V) + \Omega \frac{dI}{dt}.
\end{equation}
Barnes (2003) suggested, in agreement with results from helioseismology, 
that the relevant moment of inertia $I$, in this case is $I_*$, the moment of
inertia of the entire star. To a very good approximation, this is constant
on the main-sequence, killing the final term in Equation (16),
so that the final empirical expression for the rate of loss of angular
momentum, $dJ/dt$, from I\,sequence stars becomes
\begin{equation} \label{eqn:Jloss_Iemp}
\frac{dJ}{dt} = - \frac{\Omega^3}{8 \pi^2} I_* f^2(B-V).
\end{equation}

This expression has the appealing feature that $f(B-V)$ can be 
straight-forwardly determined from open cluster rotation period data, provided 
that the memory of the possibly complex initial conditions has been erased. 
We know this to be at least approximately true on the I\,sequence, because the 
Sun is on this sequence, and because helioseismic results indicate that the
Sun is a solid-body rotator to first order. Consequently, expression 
(17) combines a relatively well-known and easily 
calculated function $I_*$, with the observationally well-determined function 
$f(B-V)$ to state the mass dependence of angular momentum loss, 
{\it independent of assumptions beyond those encapsulated in Equations (8) and (9) above}.
On the negative side, it has the undesirable feature of mixing calculated and 
observed quantities in one expression. It is also not explanatory, because
the origin of $f(B-V)$ is not yet specified. In a section below, we propose an 
identification for $f(B-V)$.

We note that assuming $f \propto 1/\sqrt{I_*}$, as suggested
in Barnes (2003), would simply replace $I_* f^2$ with a constant, 
and would not provide the correct mass dependence observed in the rotation
period data, as shown above. Consequently, we here withdraw that suggestion.
Also, expression (17) above can be compared directly with the 
$\Omega \leq \Omega_{\rm crit}$ part of the prior loss rate expression 
(7) which is: 
\begin{equation}
\frac{dJ}{dt} = - K_w \sqrt{\frac{R}{M}} \Omega^3
\end{equation}
Comparing this with Equation (17) above shows that,
apart from numerical constants, the difference in the loss relationships
boils down to the difference between $\sqrt{R/M}$ and $I_*f^2$. 

We plot both  $I_*f^2$ and $\sqrt{R/M}$
against $B-V$ color in Figure\,6, so that a graphical comparison can be made.
As before, the curves are normalized for a solar-mass model.
Redward of $B-V = 1.4$, for very low mass stars, the physics is expected to 
be different, and neither relationship might be relevant. 
So, only the area blueward of this is relevant.
Unsurprisingly, $R/M$, and consequently, $\sqrt{R/M}$, is a relatively flat 
function of $B-V$ color for FGK stars, because the stellar radius tracks the
mass. Consequently, the angular momentum loss rate for the prior
prescription is essentially independent of spectral type, and 
{\em any assumed initial distribution of rotation periods can be expected to 
retain this initial shape when subjected to this loss rate}.
Consequently, with the prior loss prescription, the observed shape of $f(B-V)$
would have to be assumed as an initial condition. It would not arise as a
natural consequence of the angular momentum loss prescription.

\begin{figure}[h]	   
\includegraphics[scale=0.60,angle=-90]{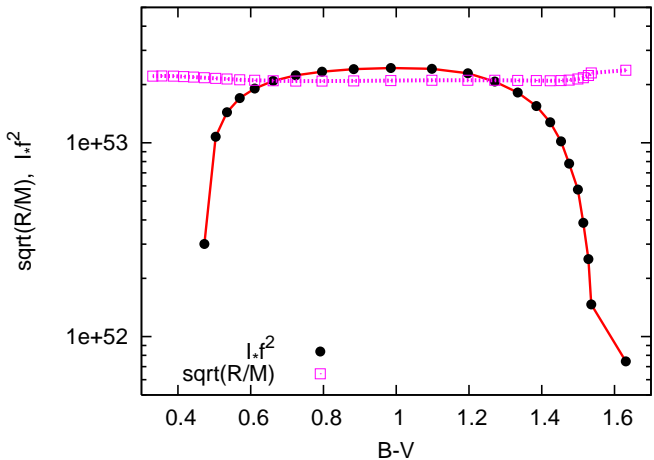}
\caption{ \small 
Comparison of the empirical dependence, $I_*f^2$ (filled circles), proposed 
here for the I\,sequence, with the prior suggested color dependence of angular 
momentum loss, $\sqrt(R/M)$, (unfilled squares). 
The curves are normalized for the solar case.
$\sqrt{R/M}$ is observed to be significantly different from the empirically 
determined function $I_* f^2$.
}
\end{figure}

In contrast, $I_*f^2$ is only somewhat flat for K stars ($B-V \sim 1$), where
it attains its maximum (Figure\,6),
and declines sharply blueward for G stars, dropping to zero blueward of late-F.
This mass-dependence has been derived from the data, so it must be true
within the uncertainties of the data themselves and the above assumptions.
Consequently the loss rate is maximized at K stars,  
but declines steadily for G stars, dropping to zero at mid-F.
Here, {\em almost any non-pathological set of initial periods would progressively attain the shape of $f(B-V)$ with the passage of time}.

\subsection{C\,sequence}

Here, we derive the second half of the angular momentum loss expression, for 
the C\,sequence, in a manner symmetrical to that for the I\,sequence above.
We can tackle the C\,sequence in a manner similar to the I\,sequence, 
although without recourse to separability of the mass and age dependencies,
by writing the rotation periods, $P_C$, of the C\,sequence stars 
as a function
\begin{equation} 
P_C = P_C(B-V,t)
\end{equation}
of the $B-V$ color and age, $t$, of the star.
The open cluster observations show that the rotation periods of the 
C\,sequence stars have a roughly exponential behavior with age\footnote{An exponential is a compact way of expressing this dependence because it can be expanded in a power series in $t/T$ and the coefficients determined. For example, for late-type stars with large $T$, the exponential will devolve into a function that is linear in $t/T$ and will also lead to the same final result for $d\Omega/dt$.}, 
suggesting that we write the periods of these stars in a form
originally suggested by Barnes (2003): 
\begin{equation}\label{eqn:P_c}
P_C = P_0 e^{t/T(B-V)}
\end{equation}
where $P_0$ is the initial period for C\,sequence stars, 
and $T(B-V)$ is the appropriate mass-dependent timescale for spin-down on the 
C\,sequence, as discussed in Section 2.2 earlier. 
The open cluster data currently provide $T(B-V)$ only in discrete form. 
Figure\,3 displays $T(B-V)$ as a function of $B-V$ color,
along with a cubic spline fit.

Switching variable from $P_C$ to $\Omega$ using $P_C = 2 \pi/ \Omega$,
equation (\ref{eqn:P_c}) becomes
\begin{equation}
 \frac{2 \pi}{\Omega} = P_0 e^{t/T(B-V)}
\end{equation}
and on differentiating it with respect to time, we get
\begin{equation}
\frac{d\Omega}{dt} = - \frac{\Omega^2}{2 \pi} P_0 \frac{e^{t/T(B-V)}}{T(B-V)}.
\end{equation}
Resubstituting from Equation (20) and again using 
$P_C = 2 \pi/ \Omega$, we get
\begin{equation}
\frac{d\Omega}{dt} = - \frac{\Omega}{T(B-V)},
\end{equation}
which again specifies the deceleration of the star in purely observational
quantities, this time for C\,sequence stars. 
Note that the mass dependence of the deceleration is 
specified completely by $1/T(B-V)$. Given the form of $T(B-V)$ from Section
2.2, with $T(B-V)$ increasing monotonically with $B-V$ color,
we see at once that as far as C\,sequence stars are concerned, the 
deceleration is greatest for late-F stars,
and declines steadily as the mass decreases through G, K, and M stars.
We note that the data provide only the deceleration above, not the angular
momentum loss rate, $dJ/dt$, derived below. However, $dJ/dt$ is required
for comparison with prior work, and is potentially important in deriving
conclusions about the mass loss rate and the Alfven Radius.

Deriving the angular momentum loss rate requires additional steps.
Applying the Chain Rule as before,
\begin{equation}
 \frac{dJ}{dt} = I_C \frac{d\Omega}{dt} + \Omega \frac{dI_C}{dt},
\end{equation}
so that substitution from above gives
\begin{equation}
 \frac{dJ}{dt} = - \frac{\Omega}{T(B-V)} I_C + \Omega \frac{dI_C}{dt}.
\end{equation}
We do not know yet what $I_C$, the relevant moment of inertia on the 
C\,sequence, is; but if we assume that it is not varying with time,
we get as the final empirical expression for the rate of loss of angular 
momentum, $dJ/dt$, from C\,sequence stars
\begin{equation} \label{eqn:Jdot_Cemp}
\frac{dJ}{dt} = - \Omega \frac{I_C}{T(B-V)}.
\end{equation}

As with the I\,sequence case, $T(B-V)$ is derivable from the data, so that
the rate of loss of angular momentum immediately follows provided that we
can guess the appropriate moment of inertia $I_C$, for C\,sequence stars,
and provided that this moment of inertia is not varying with time.
(We will revisit this last assumption in a subsequent publication.) 
As with the I\,sequence, $T(B-V)$ is empirical, and thus not explanatory,
because its origin has not been specified. In the next section, we
propose an identification for $T(B-V)$.

There is little choice about (a time-independent) $I_C$.
It could either be the moment of inertia of the entire star, $I_*$,
or, if one followed the suggestion of Barnes (2003), 
the moment of inertia of only the outer convection zone, $I_{\rm cz}$.
If $I_C = I_*$, then $dJ/dt$ would increase steeply with stellar mass, as shown 
in Figure\,7 (left panel, filled circles), because the increase in $I_*$ with 
stellar mass is amplified by the concomitant decrease in $T(B-V)$.
If $I_C = I_{\rm cz}$, then the mass dependence of $dJ/dt$ would be far less
steep with increasing stellar mass, as shown in Figure\,7 (left panel, unfilled 
circles), and undefined blueward of $B-V=0.45$, where both $I_{\rm cz}$ and 
$T(B-V)$ are zero.
(The two curves coincide for very low mass fully convective stars.)

\begin{figure}[h]     
\begin{minipage}[b]{0.5\linewidth} 
\centering
\includegraphics[width=5.75cm,angle=-90]{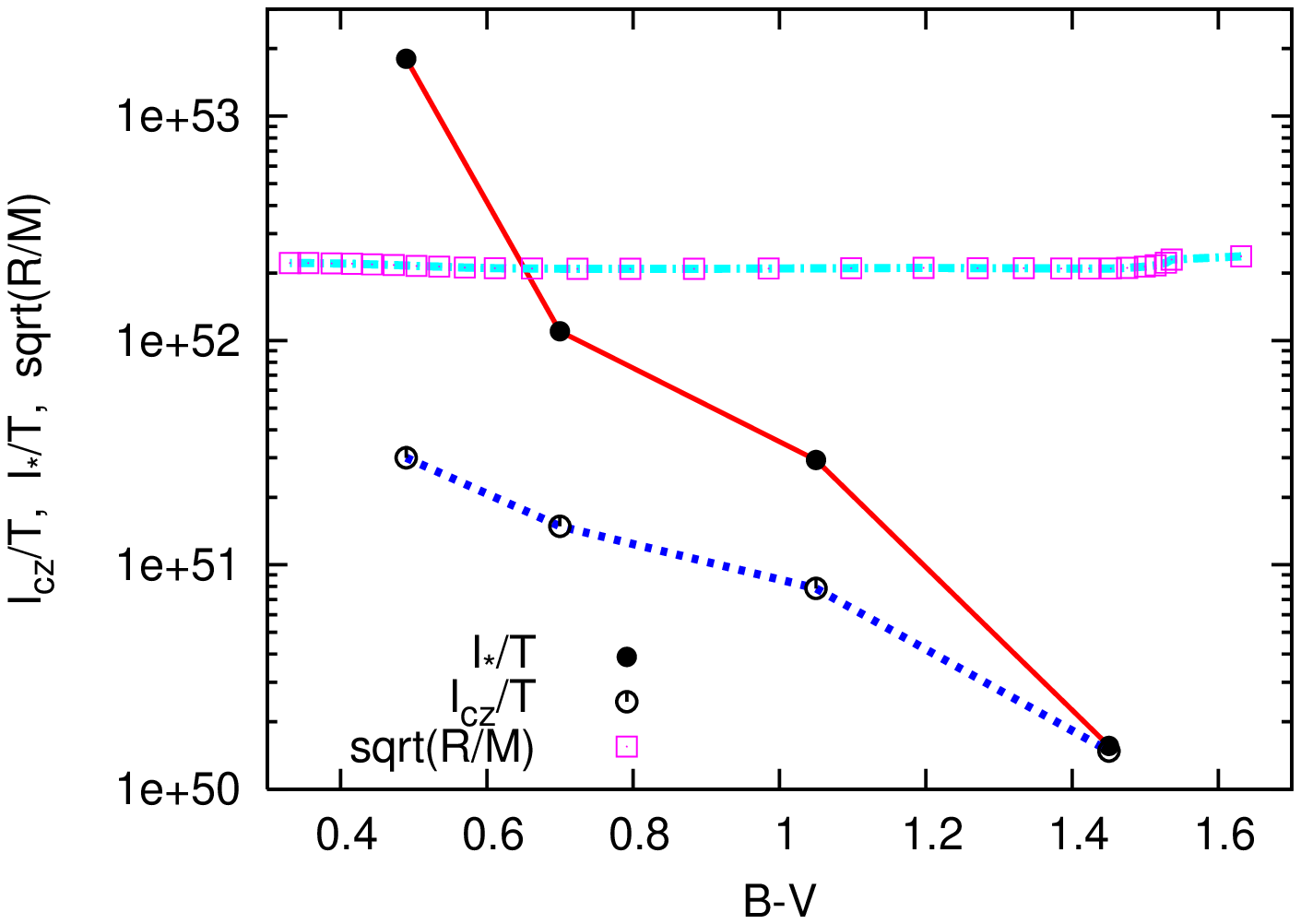}
\end{minipage}
\begin{minipage}[b]{0.5\linewidth}
\centering
\includegraphics[width=5.75cm,angle=-90]{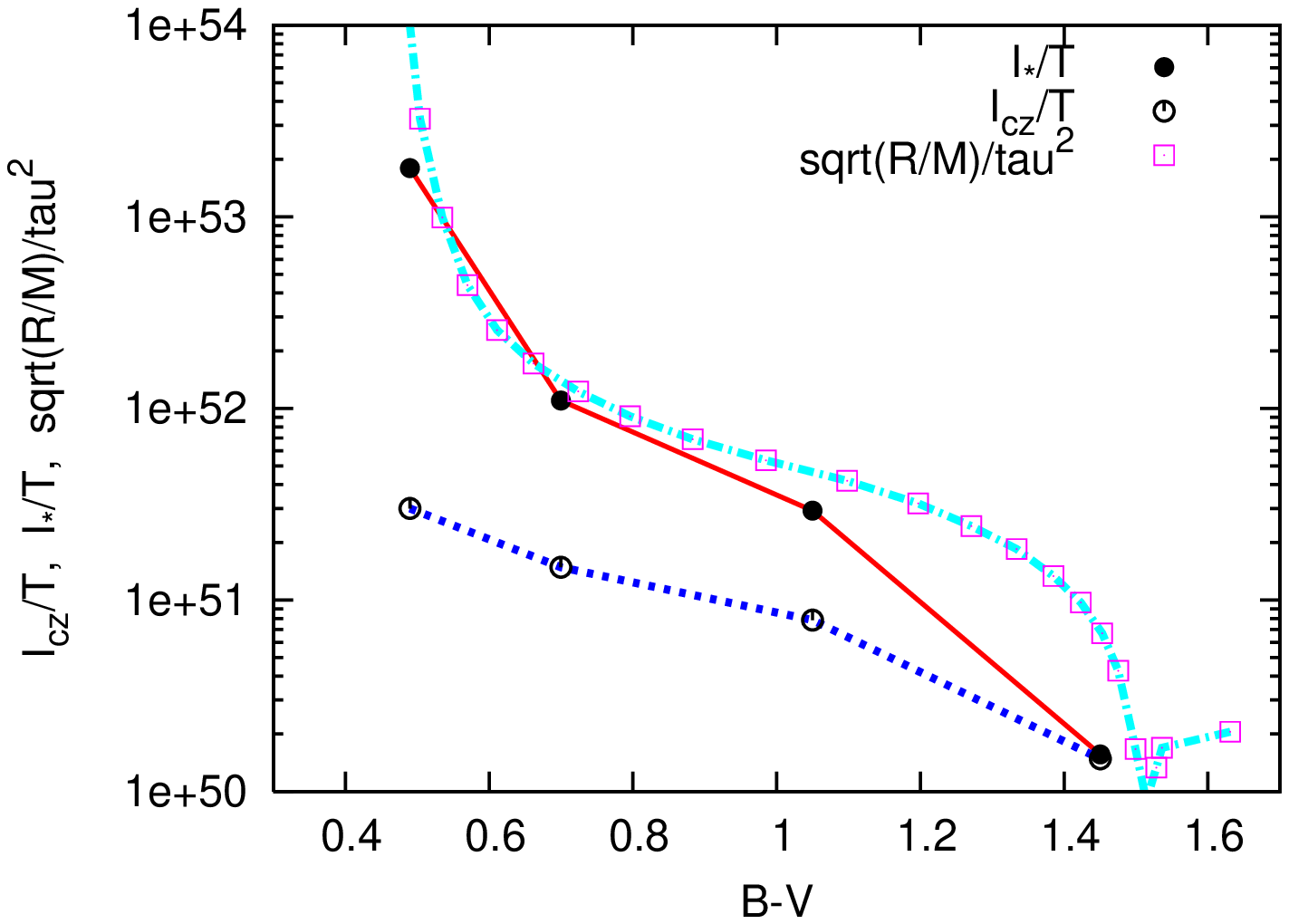}
\end{minipage}
\caption{ \small
{\bf Left:} $I_*/T(B-V)$ (filled circles) and $I_{\rm cz}/T(B-V)$ (unfilled 
circles), both candidates for the empirical C\,sequence mass dependence, are 
compared with the prior $\sqrt{R/M}$ C\,sequence mass dependence (squares).
While the first two decline steadily with increasing $B-V$ color, $\sqrt{R/M}$ 
is flat.
{\bf Right:} The same functions $I_*/T$ and $I_{\rm cz}/T$ are compared with the
Krishnamurthi et al. (1997) proposal that $\Omega_{\rm crit} \propto 1/\tau_c$ 
(instead of constant as at left).
We see that the last is similar to $I_*/T(B-V)$ and $I_{\rm cz}/T(B-V)$, 
but more complicated.
}
\end{figure}

Expression (26) can be compared directly with the 
$\Omega > \Omega_{\rm crit}$ part of the prior loss rate expression (7) which is: 
\begin{equation}
 \frac{dJ}{dt} = - K_w \sqrt{\frac{R}{M}} \Omega_{\rm crit}^2 \Omega.
\end{equation}
We see that the difference between them boils down to the difference
between $I_C/T(B-V)$ and $\sqrt{R/M} \Omega_{\rm crit}^2$. 
The case of a constant $\Omega_{\rm crit}$ is displayed graphically in Figure\,7 
(left panel, squares), normalized to the solar mass model with $I_C = I_*$. 
We note that, 
as expected, it is a remarkably flat function of stellar mass, suggesting quite 
different expectations for the corresponding angular momentum loss rate than
for the $I_C = I_{\rm cz}$ and $I_C = I_*$ cases discussed above.
Because of various issues discussed in their paper, Krishnamurthi et al. (1997)
decided to abandon a constant $\Omega_{\rm crit}$, and suggested a scaling where
$\Omega_{\rm crit} \propto 1/\tau_c$. We display the result in Figure\,7 
(right panel, squares), normalized to the solar mass model with $I_C = I_*$.
We find that this scaling gives results similar to the $I_C = I_*$ or
$I_C = I_{\rm cz}$ cases above, which are both conceptually simpler. 

\subsection{Combined expression}

To summarize the conclusions of this section, 
combining the two expressions above, 
the empirical acceleration rate becomes
\begin{equation} \label{eqn:cmb_dotomega}
\frac{d\Omega}{dt} = -
             \begin{cases} 
              {}^{\Omega} / {}_{T(B-V)} & \text{for the C\,sequence} \\
              {}^{\Omega^3\,f^2(B-V)} / {}_{8 \pi^2} & \text{for the I\,sequence},
              \end{cases}
\end{equation}
where $T(B-V)$ and $f(B-V)$ are derivable from open cluster rotation period
data. This leads immediately to the following expression for the angular
momentum loss rate
\begin{equation}   \label{eqn:cmb_gen}
\frac{dJ}{dt} = -
                \begin{cases}
                    {\scriptstyle \Omega} \, \{ {}^{I_C}/{}_{T(B-V)} -  {}^{dI_C}/{}_{dt} \} & \text{for the C\,sequence} \\
                    {}^{\Omega^3} \, {}^{I_* f^2(B-V)}/{}_{8 \pi^2} & \text{for the I\,sequence},
                 \end{cases}
\end{equation}
and on further assuming that $I_C$, the relevant moment of inertia on the 
C\,sequence is constant in time, we get
\begin{equation}   \label{eqn:cmb_spec}
\frac{dJ}{dt} = -
                \begin{cases}
                    {}^{\Omega \,I_C}/{}_{T(B-V)} & \text{for the C\,sequence} \\
                    {}^{\Omega^3 \, I_* f^2(B-V)}/{}_{8 \pi^2} & \text{for the I\,sequence.}
                 \end{cases}
\end{equation}

\section{Interpretation of $T(B-V)$ and $f(B-V)$}

We now consider, in turn, possible interpretations for $T(B-V)$ and $f(B-V)$. 
We will suggest that both are related, albeit differently, to the convective
turnover timescale in cool stars.

\subsection{$T(B-V)$}

We recall that $T(B-V)$ is defined by
\begin{equation}
 P_C = P_0 e^{t/T(B-V)},
\end{equation}
where $P_C$ represents the rotation periods of C\,sequence stars in open
clusters, $P_0$ is the appropriate initial period, and $t$ is the age.
$T(B-V)$ is the appropriate mass-dependent timescale. 
Accordingly, $T(B-V)$ must have dimensions of time.

In Section\,2.2, where it was first introduced, $T(B-V)$ was discussed in 
detail. Here, we merely reiterate that the observations currently define it in 
terms of four data points. It drops to zero at $B-V \sim$0.47, and increases 
steadily for lower mass stars, as shown in Figure\,3.

We are struck by the similarity between $T(B-V)$ and the convective
turnover timescale, $\tau_c$, in cool stars, which we have calculated and list 
in Table\,1 as a function of stellar mass and [$UBVRIJHK$] colors.
$\tau_c$ and $T(B-V)$ are displayed and compared in Figure\,8 (left) and 
Figure\,8 (right) on linear and logarithmic
scales respectively. In this comparison, and the next, we use the global
turnover timescale. However, the (somewhat shorter) local timescale at the base 
of the convection zone is also provided in Table\,1.
More details about the calculation can be found in Kim \& Demarque (1996).

\begin{figure}[h]     
\begin{minipage}[b]{0.5\linewidth} 
\centering
\includegraphics[width=5.75cm,angle=-90]{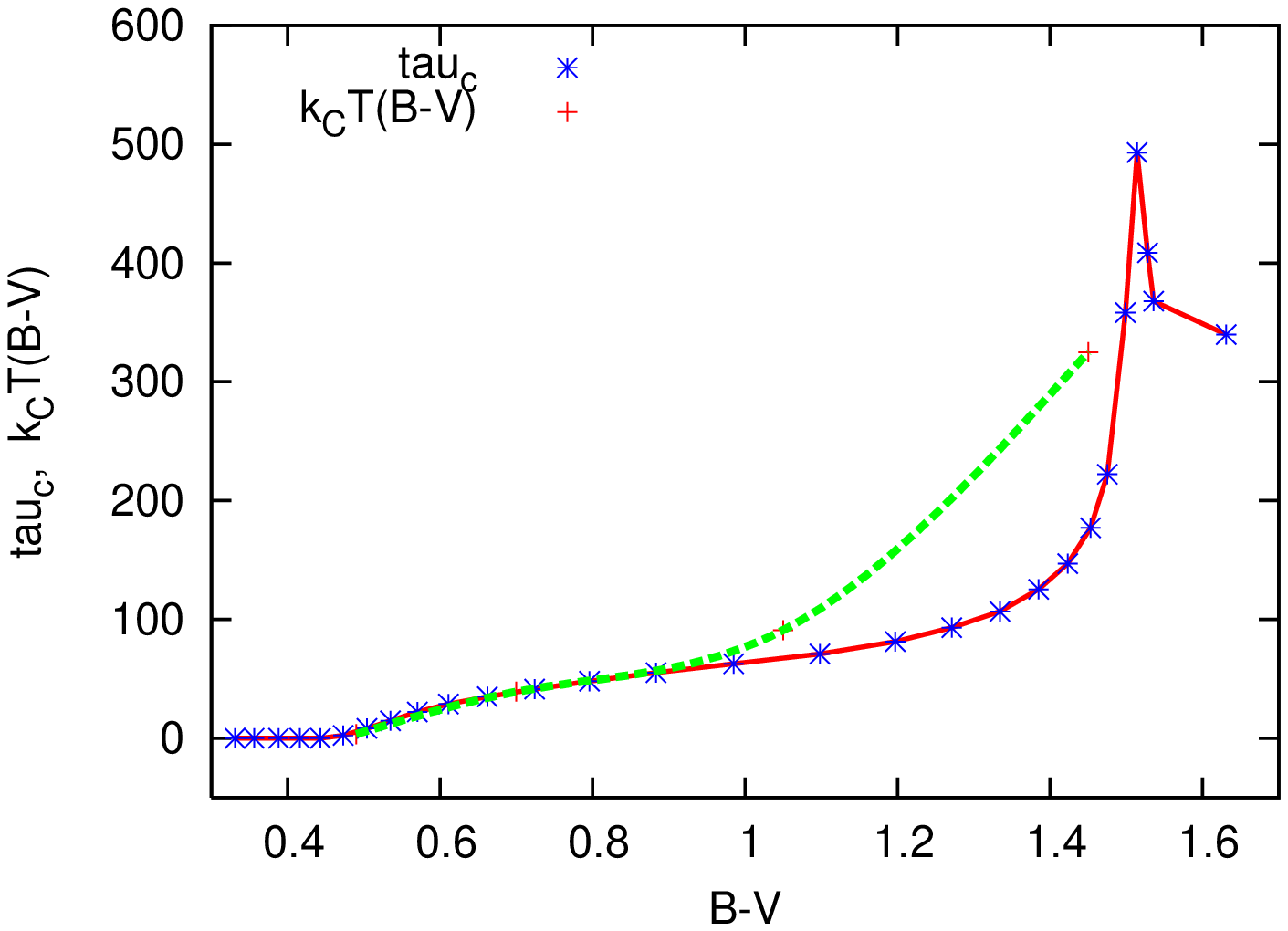}
\end{minipage}
\begin{minipage}[b]{0.5\linewidth}
\centering
\includegraphics[width=5.75cm,angle=-90]{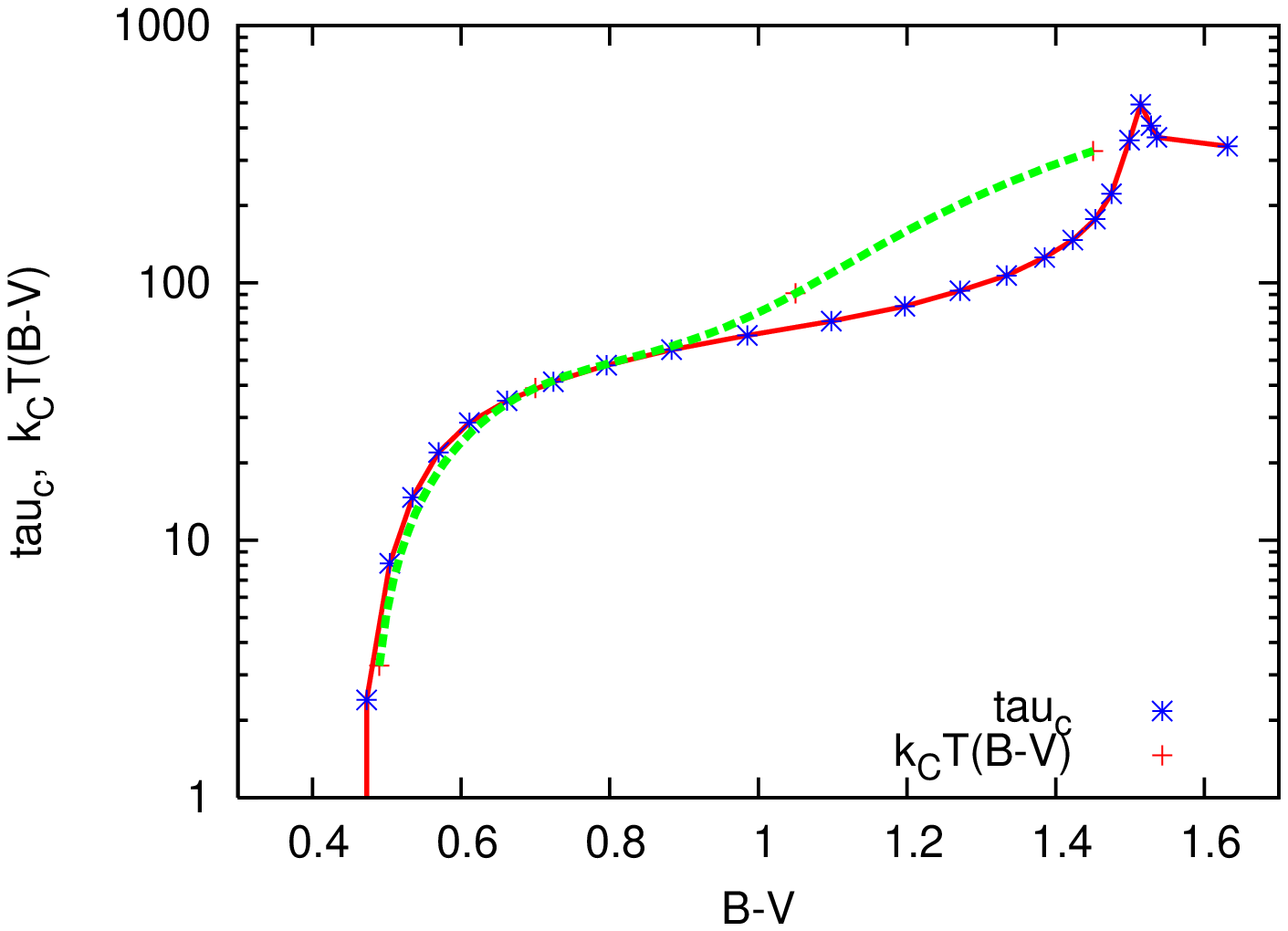}
\end{minipage}
\caption{ \small 
{\bf Left:} $\tau_c$ (asterisks) and $k_C T(B-V)$ (pluses) are plotted against 
$B-V$ color on a linear scale. Normalization at solar mass requires 
$k_C = 0.65$d/Myr. A cubic spline fit to the $T(B-V)$ data points (dashed green 
line) is also displayed.
{\bf Right:} The same comparison on a logarithmic scale, emphasizing the blue 
region where both functions decline sharply.
We see that $\tau_c$ and $k_C T(B-V)$ are in reasonable agreement.
}
\end{figure}

The behaviors of $\tau_c$ and $T(B-V)$ suggest the identification
\begin{equation}
T(B-V) =  \frac{\tau_c}{k_C},
\end{equation}
where $\tau_c$ is the convective turnover timescale, and $k_C$ is a 
(dimensionless) constant. We find that $k_C \approx 0.65$ d/Myr.
Indeed, with this choice of $k_C$, the two curves are in reasonable agreement.
(At this early stage, it is not possible to comment on the detectability or
importance of the slight non-monotonicity of $\tau_c$ for the lowest stellar 
masses.)


\subsection{$f(B-V)$}

The situation with $f(B-V)$ is similar to that with $T(B-V)$ above.
Recall that $f(B-V)$ is defined by the expression
\begin{equation}
P_I = f(B-V) \times g(t),
\end{equation}
where $P_I$ represents the rotation periods of I\,sequence stars in open
clusters, and $f(B-V)$ and $g(t)$ are separable functions, respectively, 
of the $B-V$ color and age, $t$, of the stars. 
To a precision adequate for the present purposes, $g(t) = \sqrt{t}$. 
Because $P_I$ has the dimensions of $time$
we infer that $f(B-V)$ must have the dimensions of $\sqrt{time}$, 
or that $f^2(B-V)$ must have the dimensions of $time$.

The rotation periods of the I\,sequence stars decrease sharply for stars bluer
than the Sun, approaching zero for $B-V = 0.45$. For stars redder than the Sun, 
the rotation periods increase all the way till at least 
$0.5 M_{\odot}$. This behavior is again remarkably reminiscent of the
behavior of the convective turnover timescale but dimensionally ought to
be associated with its square root. We therefore propose the identification
\begin{equation}
f^2(B-V) = \frac{2 \tau_c}{k_I},
\end{equation}
where $k_I$ is a dimensionless constant, and $\tau_c$ is again the convective
turnover timescale. The factor of $2$ is added purely for later convenience.
We find that $k_I \sim 740$ or $1340$ Myr/d, where the smaller value is relevant
to the $f(B-V)$ determination of Meibom et al. (2009)
while the larger value is 
relevant to the determination of Mamajek \& Hillenbrand (2008). 
The functions $2 \tau_c$ and $k_I f^2(B-V)$ are compared graphically
in Figure\,9 and indeed they are in reasonable agreement.

\begin{figure}[h]     
\begin{minipage}[b]{0.5\linewidth} 
\centering
\includegraphics[width=5.75cm,angle=-90]{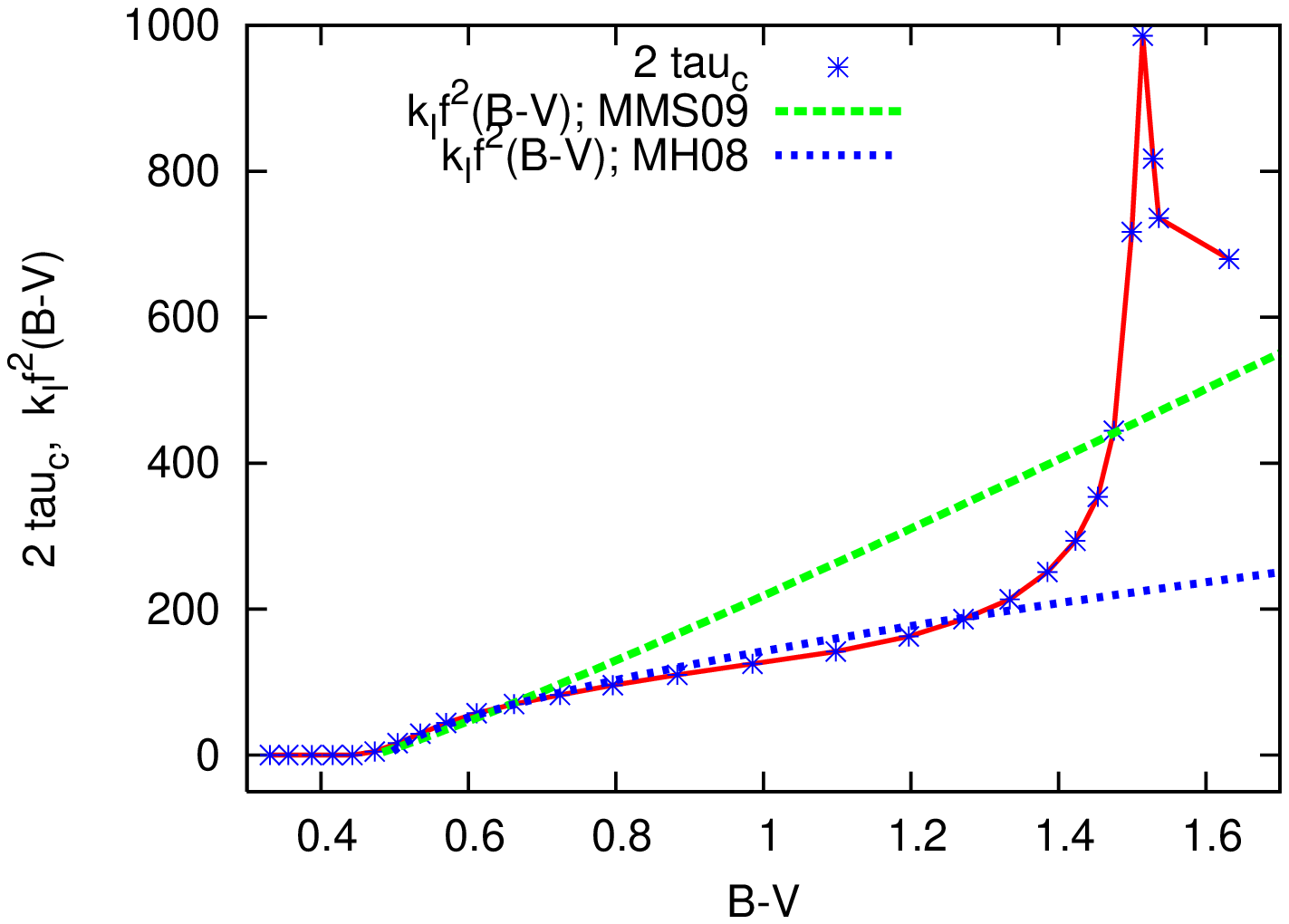}
\end{minipage}
\begin{minipage}[b]{0.5\linewidth}
\centering
\includegraphics[width=5.75cm,angle=-90]{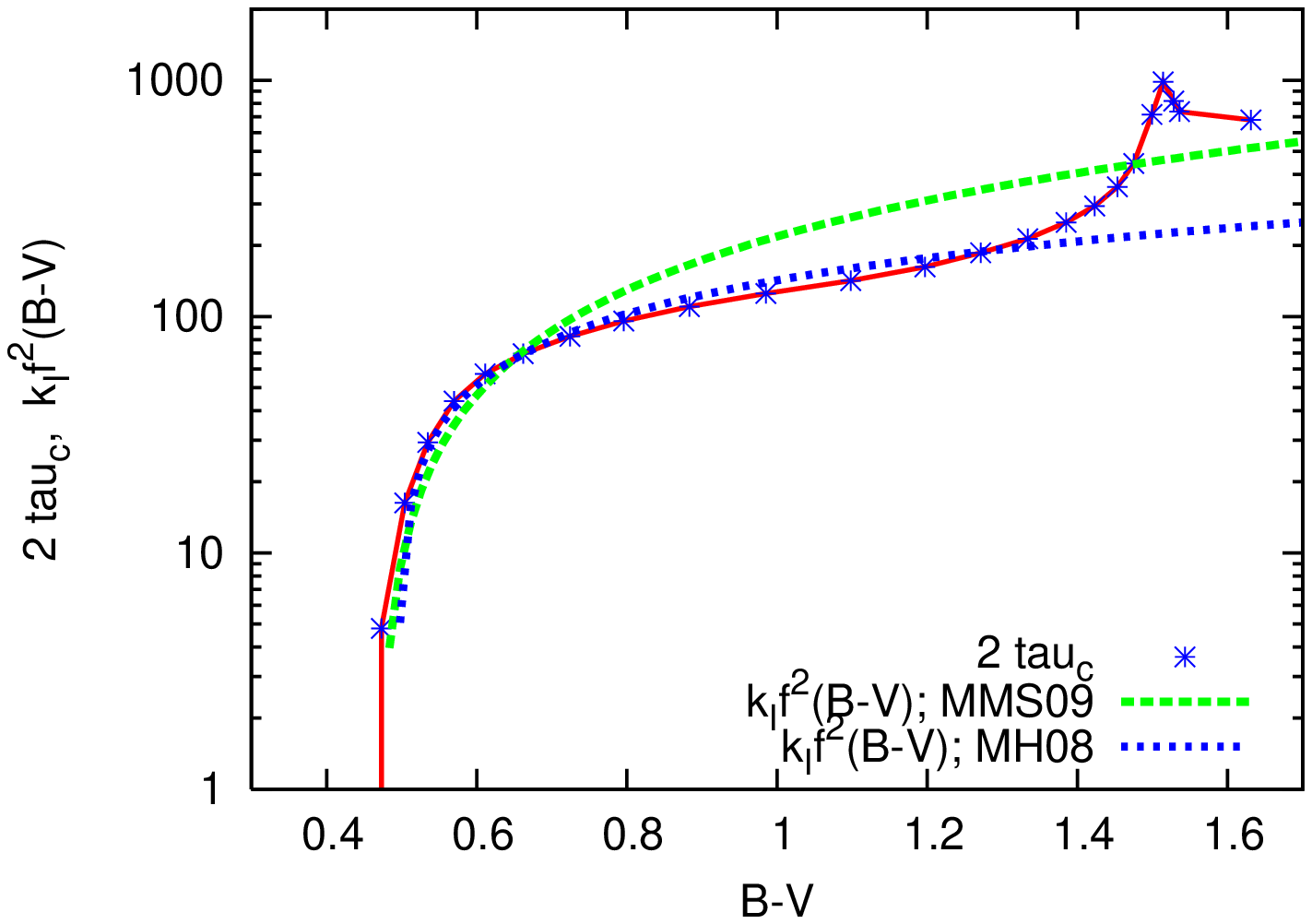}
\end{minipage}
\caption{ \small 
{\bf Left:} $2 \tau_c$ (asterisks) and $k_I f^2(B-V)$ are plotted against $B-V$ 
color on a linear scale. The dashed green curve is the Meibom et al. (2009)
determination of $f(B-V)$, while the dotted blue curve is that of 
Mamajek \& Hillenbrand (2008).
Normalization at solar mass requires $k_I = 740$Myr/d and 
$1340$Myr/d respectively.
{\bf Right:} The same comparison on a logarithmic scale, emphasizing the blue 
region where all the functions decline sharply.
We see that $2 \tau_c$ and $k_I f^2(B-V)$ are in reasonable agreement.
}
\end{figure}

\subsection{Symmetric combination}

The above interpretations of $T(B-V)$ and of $f(B-V)$ are appealingly
symmetric. Although there are two timescales in the problem, captured
by the dimensionless constants, $k_C$ and $k_I$, the spin-down of both
the C- and I\,sequences in open clusters
appear to be connected by one underlying variable, $\tau_c$, the convective
turnover timescale in cool stars.
As a result, it appears that the underlying variables in stellar
rotation appear to be $P$, $t$, and $\tau_c$, the last encapsulating
the dependence on stellar mass.

Consequently, placing the identifications above into Equation (28), we suggest the 
following expression for the deceleration of cool stars:
\begin{equation} \label{eqn:cmb_dotomega_fin}
\frac{d\Omega}{dt} = -
                \begin{cases}
                {}^{\Omega \,k_C}/{}_{\tau_c} & \text{for the C\,sequence} \\
                {}^{\Omega^3 \, \tau_c}/{}_{4 \pi^2 k_I} & \text{for the I\,sequence}
                 \end{cases}
\end{equation}
or, on changing variable from $\Omega$ to $P = 2\pi/\Omega$, we get
\begin{equation} \label{eqn:cmb_dotP}
\frac{dP}{dt} = 
                \begin{cases}
                 {}^{k_C P}/{}_{\tau_c}  & \text{for the C\,sequence} \\
                 {}^{\tau_c}/{}_{k_I P} & \text{for the I\,sequence} 
                 \end{cases}
\end{equation}
which displays the symmetry of the problem explicitly. 

\section{Relationship with stellar activity physics}

In a classic paper, Noyes et al. (1984) proposed that the fractional
chromospheric emission from cool stars was dependent, not just on the
rotation period, $P$, but also on stellar mass. 
Inspired by theoretical work by Durney \& Latour (1978), they identified the
convective turnover timescale, $\tau_c(B-V)$, as the most likely 
variable to capture the dependence on stellar mass, and demonstrated
that the fractional chromospheric emission of cool stars, $R'_{HK}$, 
was more tightly dependent on the Rossby Number, $Ro = P/\tau_c$, 
than on $P$ alone.
Soon thereafter, the transition-region ultraviolet emission and coronal 
X-Ray activity of cool stars were also shown to behave similarly 
(Vilhu 1984; Simon et al. 1985), and it is now generally accepted that
stellar activity is associated more closely with the Rossby Number
than with the rotation period alone.

This work brings the rotational evolution of cool stars under the same
umbrella by showing that $dP/dt$ is a function of $P/\tau_c$.
Consequently, the rotation rate of a cool star appears to be a deterministic
function of its age and its mass, the latter appearing through the
convective turnover timescale, $\tau_c$.

\section{Conclusions}

We have shown here that open cluster rotation period data allow the empirical
determination of an expression for angular momentum loss from cool stars on
the main-sequence. 
The relationship has a bifurcation, as before, that corresponds to an observed
bifurcation in the rotation periods of open cluster stars.

One component of the relationship, the dependence on rotation rate, persists
from prior work. The remaining dependence is captured by two observationally
determinable functions of the mass or color of a star, which we write as
$f(B-V)$ and $T(B-V)$ for the two tines of the relationship, respectively.
Transformations to mass or other desired colors in the set [$UBVRIJHK$] can be 
accomplished using Table\,1.

We derive an empirical expression for the angular momentum loss rate, which is
\begin{equation}
\frac{dJ}{dt} = -
                \begin{cases}
                   {\scriptstyle \Omega} \, \{ {}^{I_C}/{}_{T} -  {}^{dI_C}/{}_{dt} \} & \text{for the C\,sequence} \\
                   {}^{\Omega^3} \, {}^{I_* f^2}/{}_{8 \pi^2} & \text{for the I\,sequence} 
                 \end{cases}
\end{equation}
and which simplifies to 
\begin{equation}
\frac{dJ}{dt} = -
                \begin{cases}
                {}^{\Omega \,I_C}/{}_T  & \text{for the C\,sequence} \\
                {}^{\Omega^3 \, I_* f^2}/{}_{8 \pi^2} & \text{for the I\,sequence.}
                 \end{cases}
\end{equation}
if the relevant moment of inertia, $I_C$, on the C\,sequence, is not 
time-varying.

Both $f(B-V)$ and $T(B-V)$ appear to be related to the convective turnover
timescale, $\tau_c$ in stars. We suggest the identifications 
\begin{equation}
T(B-V) = \frac{\tau_c}{k_C}
\end{equation} 
and 
\begin{equation}
f^2(B-V)= \frac{2 \tau_c}{k_I},
\end{equation} 
where $k_C$ and $k_I$ are two dimensionless constants appropriate to the 
C- and I\,sequences, and are respectively approximately
0.65\,d/Myr and 740 or 1340\,Myr/d, 
depending on which of two particular forms one uses for $f(B-V)$.

Consequently our final expression for the angular momentum loss rate is
\begin{equation}  
\frac{dJ}{dt} = -
                \begin{cases}
        {\scriptstyle \Omega} \, \{ {}^{k_C I_C}/{}_{\tau_c} -  {}^{dI_C}/{}_{dt} \} & \text{for the C\,sequence} \\
       {}^{\Omega^3 \, I_*\tau_c}/{}_{4 \pi^2 k_I} & \text{for the I\,sequence}
                  \end{cases}
\end{equation}
and that for the deceleration of cool stars is
\begin{equation} 
\frac{dP}{dt} = 
                \begin{cases} 
                 {}^{k_C P} / {}_{\tau_c} & \text{for the C\,sequence} \\
                 {}^{\tau_c} / {}_{k_I P} & \text{for the I\,sequence}
                 \end{cases}
\end{equation}
which is both symmetric and dimensionally correct.
We see that in this model, the evolution of the rotation period, $P$,
depends only on the age, $t$, and the convective turnover timescale, $\tau_c$,
which encodes the mass dependence of rotation.

Finally, we have pointed out that this model for the rotational evolution of 
stars makes a natural connection to stellar activity physics, where $P/\tau_c$
is the preferred independent variable.

These ideas are developed further in following papers into a simple 
nonlinear model for the rotational evolution of cool stars from C- to I-type, 
and a companion paper will provide more general grids of related calculated 
quantities on and off the main-sequence.

\acknowledgements

S.A.B. acknowledges support for this work from NASA through Spitzer award GO5 
50465 to Lowell Observatory.
Y.-C.K. is grateful to the Yon Am Foundation for its generous financial support
during his sabbatical year at Lowell Observatory, and to Northern Arizona
University for help with related visa issues.
The authors appreciate the prompt and constructive critique from the referee, 
Alexander Scholz.


\newpage
 

\begin{thebibliography}

\bibitem[{{Andronov} {et~al.}(2003){Andronov}, {Pinsonneault}, \&
  {Sills}}]{And+03}
{Andronov}, N., {Pinsonneault}, M., \& {Sills}, A. 2003, \apj, 582, 358

\bibitem[{{Armitage} \& {Clarke}(1996)}]{AC96}
{Armitage}, P.~J., \& {Clarke}, C.~J. 1996, \mnras, 280, 458

\bibitem[{{Barnes} \& {Sofia}(1996)}]{BS96}
{Barnes}, S., \& {Sofia}, S. 1996, \apj, 462, 746

\bibitem[{{Barnes} {et~al.}(2001){Barnes}, {Sofia}, \& {Pinsonneault}}]{BSP01}
{Barnes}, S., {Sofia}, S., \& {Pinsonneault}, M. 2001, \apj, 548, 1071

\bibitem[{{Barnes}(2003)}]{Bar03a}
{Barnes}, S.~A. 2003, \apj, 586, 464

\bibitem[{{Barnes}(2007)}]{Bar07}
---. 2007, \apj, 669, 1167

\bibitem[{{Belcher} \& {MacGregor}(1976)}]{BM76}
{Belcher}, J.~W., \& {MacGregor}, K.~B. 1976, \apj, 210, 498

\bibitem[{{Bouvier} {et~al.}(1997){Bouvier}, {Forestini}, \& {Allain}}]{BFA97}
{Bouvier}, J., {Forestini}, M., \& {Allain}, S. 1997, \aap, 326, 1023

\bibitem[{{Chaboyer} {et~al.}(1995){Chaboyer}, {Demarque}, \&
  {Pinsonneault}}]{Cha+95}
{Chaboyer}, B., {Demarque}, P., \& {Pinsonneault}, M.~H. 1995, \apj, 441, 876

\bibitem[{{Collier Cameron} \& {Campbell}(1993)}]{CC93}
{Collier Cameron}, A., \& {Campbell}, C.~G. 1993, \aap, 274, 309


\bibitem[{{Collier Cameron} {et~al.}(2009){Collier Cameron}, {Davidson},
  {Hebb}, {Skinner}, {Anderson}, {Christian}, {Clarkson}, {Enoch}, {Irwin},
  {Joshi}, {Haswell}, {Hellier}, {Horne}, {Kane}, {Lister}, {Maxted}, {Norton},
  {Parley}, {Pollacco}, {Ryans}, {Scholz}, {Skillen}, {Smalley}, {Street},
  {West}, {Wilson}, \& {Wheatley}}]{CC+_comaber}
{Collier Cameron}, A., {Davidson}, V.~A., {Hebb}, L., {Skinner}, G.,
  {Anderson}, D.~R., {Christian}, D.~J., {Clarkson}, W.~I., {Enoch}, B.,
  {Irwin}, J., {Joshi}, Y., {Haswell}, C.~A., {Hellier}, C., {Horne}, K.~D.,
  {Kane}, S.~R., {Lister}, T.~A., {Maxted}, P.~F.~L., {Norton}, A.~J.,
  {Parley}, N., {Pollacco}, D., {Ryans}, R., {Scholz}, A., {Skillen}, I.,
  {Smalley}, B., {Street}, R.~A., {West}, R.~G., {Wilson}, D.~M., \&
  {Wheatley}, P.~J. 2009, \mnras, 400, 451

\bibitem[{{Collier Cameron} \& {Li}(1994)}]{CCL94}
{Collier Cameron}, A., \& {Li}, J. 1994, \mnras, 269, 1099

\bibitem[{{Denissenkov} {et~al.}(2010){Denissenkov}, {Pinsonneault},
  {Terndrup}, \& {Newsham}}]{Den+10}
{Denissenkov}, P.~A., {Pinsonneault}, M., {Terndrup}, D.~M., \& {Newsham}, G.
  2010, \apj, 716, 1269

\bibitem[{{Durney} \& {Latour}(1978)}]{DL78}
{Durney}, B.~R., \& {Latour}, J. 1978, Geophysical and Astrophysical Fluid
  Dynamics, 9, 241

\bibitem[{{Endal} \& {Sofia}(1981)}]{ES81}
{Endal}, A.~S., \& {Sofia}, S. 1981, \apj, 243, 625

\bibitem[{{Green} {et~al.}(1987){Green}, {Demarque}, \& {King}}]{GDK87}
{Green}, E.~M., {Demarque}, P., \& {King}, C.~R. 1987, {The revised Yale
  isochrones and luminosity functions} (Yale University Observatory, New Haven)

\bibitem[{{Hartman} {et~al.}(2010){Hartman}, {Bakos}, {Kov{\'a}cs}, \&
  {Noyes}}]{Har+_pleiades}
{Hartman}, J.~D., {Bakos}, G.~{\'A}., {Kov{\'a}cs}, G., \& {Noyes}, R.~W. 2010,
  \mnras, 1162

\bibitem[{{Hartman} {et~al.}(2009){Hartman}, {Gaudi}, {Pinsonneault}, {Stanek},
  {Holman}, {McLeod}, {Meibom}, {Barranco}, \& {Kalirai}}]{Har+_m37}
{Hartman}, J.~D., {Gaudi}, B.~S., {Pinsonneault}, M.~H., {Stanek}, K.~Z.,
  {Holman}, M.~J., {McLeod}, B.~A., {Meibom}, S., {Barranco}, J.~A., \&
  {Kalirai}, J.~S. 2009, \apj, 691, 342

\bibitem[{{Irwin} {et~al.}(2009){Irwin}, {Aigrain}, {Bouvier}, {Hebb},
  {Hodgkin}, {Irwin}, \& {Moraux}}]{Irw+_M50}
{Irwin}, J., {Aigrain}, S., {Bouvier}, J., {Hebb}, L., {Hodgkin}, S., {Irwin},
  M., \& {Moraux}, E. 2009, \mnras, 392, 1456

\bibitem[{{James} {et~al.}(2010){James}, {Barnes}, {Meibom}, {Lockwood},
  {Levine}, {Deliyannis}, {Platais}, {Steinhauer}, \& {Hurley}}]{Jam+_m34}
{James}, D.~J., {Barnes}, S.~A., {Meibom}, S., {Lockwood}, G.~W., {Levine},
  S.~E., {Deliyannis}, C., {Platais}, I., {Steinhauer}, A., \& {Hurley}, B.~K.
  2010, \aap, 515, A100+

\bibitem[{{Kawaler}(1988)}]{Kaw88}
{Kawaler}, S.~D. 1988, \apj, 333, 236

\bibitem[{{Kawaler}(1989)}]{Kaw89}
---. 1989, \apjl, 343, L65

\bibitem[{{Keppens} {et~al.}(1995){Keppens}, {MacGregor}, \&
  {Charbonneau}}]{Kep+95}
{Keppens}, R., {MacGregor}, K.~B., \& {Charbonneau}, P. 1995, \aap, 294, 469

\bibitem[{{Kim} \& {Demarque}(1996)}]{KD96}
{Kim}, Y., \& {Demarque}, P. 1996, \apj, 457, 340

\bibitem[{{Kraft}(1967)}]{Kra67}
{Kraft}, R.~P. 1967, \apj, 150, 551

\bibitem[{{Krishnamurthi} {et~al.}(1997){Krishnamurthi}, {Pinsonneault},
  {Barnes}, \& {Sofia}}]{KPBS97}
{Krishnamurthi}, A., {Pinsonneault}, M.~H., {Barnes}, S., \& {Sofia}, S. 1997,
  \apj, 480, 303

\bibitem[{{Lejeune} {et~al.}(1997){Lejeune}, {Cuisinier}, \& {Buser}}]{LCB97}
{Lejeune}, T., {Cuisinier}, F., \& {Buser}, R. 1997, \aaps, 125, 229

\bibitem[{{Lejeune} {et~al.}(1998){Lejeune}, {Cuisinier}, \& {Buser}}]{LCB98}
---. 1998, \aaps, 130, 65

\bibitem[{{MacGregor} \& {Brenner}(1991)}]{MB91}
{MacGregor}, K.~B., \& {Brenner}, M. 1991, \apj, 376, 204

\bibitem[{{Maeder}(2009)}]{Mae09}
{Maeder}, A. 2009, {Physics, Formation and Evolution of Rotating Stars}
  (Springer: Berlin, Heidelberg)

\bibitem[{{Mamajek} \& {Hillenbrand}(2008)}]{MH08}
{Mamajek}, E.~E., \& {Hillenbrand}, L.~A. 2008, \apj, 687, 1264

\bibitem[{{Matt} \& {Pudritz}(2008)}]{MP08}
{Matt}, S., \& {Pudritz}, R.~E. 2008, \apj, 678, 1109

\bibitem[{{Meibom} {et~al.}(2006){Meibom}, {Mathieu}, \& {Stassun}}]{MMS06}
{Meibom}, S., {Mathieu}, R.~D., \& {Stassun}, K.~G. 2006, \apj, 653, 621

\bibitem[{{Meibom} {et~al.}(2009){Meibom}, {Mathieu}, \& {Stassun}}]{MMS_m35}
---. 2009, \apj, 695, 679

\bibitem[{{Mestel}(1968)}]{Mes68}
{Mestel}, L. 1968, \mnras, 138, 359

\bibitem[{{Mestel}(1984)}]{Mes84}
{Mestel}, L. 1984, in Lecture Notes in Physics, Berlin Springer Verlag, Vol.
  193, Cool Stars, Stellar Systems, and the Sun, ed. S.~L. {Baliunas} \&
  L.~{Hartmann}, 49--+

\bibitem[{{Noyes} {et~al.}(1984){Noyes}, {Hartmann}, {Baliunas}, {Duncan}, \&
  {Vaughan}}]{Noy+84}
{Noyes}, R.~W., {Hartmann}, L.~W., {Baliunas}, S.~L., {Duncan}, D.~K., \&
  {Vaughan}, A.~H. 1984, \apj, 279, 763

\bibitem[{{Parker}(1958)}]{Par58}
{Parker}, E.~N. 1958, \apj, 128, 664

\bibitem[{{Pinsonneault} {et~al.}(1989){Pinsonneault}, {Kawaler}, {Sofia}, \&
  {Demarque}}]{PKSD89}
{Pinsonneault}, M.~H., {Kawaler}, S.~D., {Sofia}, S., \& {Demarque}, P. 1989,
  \apj, 338, 424

\bibitem[{{Rappaport} {et~al.}(1983){Rappaport}, {Verbunt}, \& {Joss}}]{Rap+83}
{Rappaport}, S., {Verbunt}, F., \& {Joss}, P.~C. 1983, \apj, 275, 713

\bibitem[{{Schatzman}(1962)}]{Sch62}
{Schatzman}, E. 1962, Annales d'Astrophysique, 25, 18

\bibitem[{{Scholz} \& {Eisl{\"o}ffel}(2007)}]{SE_praesepe}
{Scholz}, A., \& {Eisl{\"o}ffel}, J. 2007, \mnras, 381, 1638

\bibitem[{{Scholz} {et~al.}(2009){Scholz}, {Eisl{\"o}ffel}, \&
  {Mundt}}]{SEM_4665}
{Scholz}, A., {Eisl{\"o}ffel}, J., \& {Mundt}, R. 2009, \mnras, 400, 1548

\bibitem[{{Sills} {et~al.}(2000){Sills}, {Pinsonneault}, \& {Terndrup}}]{SPT00}
{Sills}, A., {Pinsonneault}, M.~H., \& {Terndrup}, D.~M. 2000, \apj, 534, 335

\bibitem[{{Simon} {et~al.}(1985){Simon}, {Herbig}, \& {Boesgaard}}]{Sim+85}
{Simon}, T., {Herbig}, G., \& {Boesgaard}, A.~M. 1985, \apj, 293, 551

\bibitem[{{Skumanich}(1972)}]{Sku72}
{Skumanich}, A. 1972, \apj, 171, 565

\bibitem[{{Stauffer}(1994)}]{Sta94}
{Stauffer}, J. 1994, in Astronomical Society of the Pacific Conference Series,
  Vol.~64, Cool Stars, Stellar Systems, and the Sun, ed. {J.-P.~Caillault},
  163--+

\bibitem[{{van Leeuwen} \& {Alphenaar}(1982)}]{vA82}
{van Leeuwen}, F., \& {Alphenaar}, P. 1982, The Messenger, 28, 15

\bibitem[{{Vilhu}(1984)}]{Vil84}
{Vilhu}, O. 1984, \aap, 133, 117

\bibitem[{{Weber} \& {Davis}(1967)}]{WD67}
{Weber}, E.~J., \& {Davis}, L.~J. 1967, \apj, 148, 217

\end{thebibliography}


\clearpage
\begin{landscape}
\begin{deluxetable}{rrrrrrrrrrrrrrrrrrrrrrr}
\tabletypesize{\tiny}
\setlength{\tabcolsep}{0.02in}
\tablecolumns{23}
\tablewidth{0pc}
\tablecaption{Calculated Properties of Solar Metallicity Main-sequence Stellar Models at 500\,Myr, as used in the text}
\tablehead{
\colhead{Mass} & \colhead{log\,T} & \colhead{log\,$L/L_{\odot}$} & 
\colhead{Age} & \colhead{Global $\tau_c$} & \colhead{Local $\tau_c$} &
\multicolumn{4}{c}{Moment of Inertia ($g\,cm^2$)} &
\multicolumn{8}{c}{Lejeune et al. (1997, 1998) Colors} & 
\multicolumn{5}{c}{Green et al. (1987) Colors} \\
\colhead{($M_{\odot}$)} & \colhead{(K)}   & \colhead{-}  & \colhead{(Gyr)} &
\colhead{(d)}    & \colhead{(d)}   & 
\colhead{Con.\,Core} & \colhead{Rad.\,Zone} & \colhead{Con.\,Env} & 
\colhead{Total} & 
\colhead{U} & \colhead{B} & \colhead{V} & \colhead{R} & \colhead{I} &
\colhead{J} & \colhead{H} & \colhead{K} &
\colhead{U} & \colhead{B} & \colhead{V} & \colhead{R} & \colhead{I} }
\startdata
0.15 & 3.50961 & -2.53866 & 0.500 & 3.398e+02 & 1.628e+02 & 0.000e+00 & 0.000e+00 & 8.922e+51 & 8.922e+51 & 16.391 & 15.056 & 13.425 & 12.221 & 10.701 & 9.254 & 8.640 & 8.385 & 16.110 & 14.982 & 13.420 & 12.265 & 10.819\\ 
0.20 & 3.52474 & -2.28585 & 0.500 & 3.679e+02 & 1.784e+02 & 0.000e+00 & 0.000e+00 & 1.839e+52 & 1.839e+52 & 15.172 & 14.000 & 12.464 & 11.357 &  9.986 & 8.666 & 8.041 & 7.805 & 15.259 & 14.095 & 12.563 & 11.471 & 10.129\\ 
0.25 & 3.53505 & -2.10378 & 0.500 & 4.086e+02 & 2.030e+02 & 0.000e+00 & 0.000e+00 & 3.166e+52 & 3.166e+52 & 14.472 & 13.316 & 11.788 & 10.719 &  9.484 & 8.282 & 7.652 & 7.432 & 14.575 & 13.434 & 11.939 & 10.896 &  9.634\\ 
0.30 & 3.54339 & -1.95807 & 0.500 & 4.930e+02 & 2.565e+02 & 0.000e+00 & 0.000e+00 & 4.907e+52 & 4.907e+52 & 13.896 & 12.761 & 11.247 & 10.214 &  9.089 & 7.976 & 7.339 & 7.133 & 14.069 & 12.916 & 11.442 & 10.436 &  9.238\\ 
0.35 & 3.55082 & -1.81901 & 0.500 & 3.584e+02 & 1.718e+02 & 1.330e+51 & 2.140e+51 & 6.991e+52 & 7.338e+52 & 13.457 & 12.308 & 10.809 &  9.802 &  8.729 & 7.641 & 7.006 & 6.806 & 13.596 & 12.440 & 10.986 & 10.004 &  8.864\\ 
0.40 & 3.55929 & -1.69122 & 0.500 & 2.221e+02 & 1.133e+02 & 7.920e+50 & 2.080e+52 & 7.952e+52 & 1.011e+53 & 13.042 & 11.873 & 10.398 &  9.422 &  8.402 & 7.332 & 6.694 & 6.503 & 13.163 & 11.983 & 10.545 &  9.588 &  8.515\\ 
0.45 & 3.56950 & -1.55867 & 0.500 & 1.769e+02 & 9.196e+01 & 7.784e+50 & 4.435e+52 & 8.833e+52 & 1.334e+53 & 12.600 & 11.402 &  9.949 &  9.008 &  8.056 & 7.014 & 6.377 & 6.195 & 12.653 & 11.470 & 10.061 &  9.135 &  8.148\\ 
0.50 & 3.58154 & -1.42169 & 0.500 & 1.468e+02 & 7.728e+01 & 6.085e+50 & 7.457e+52 & 9.507e+52 & 1.702e+53 & 12.117 & 10.895 &  9.472 &  8.566 &  7.680 & 6.705 & 6.065 & 5.897 & 12.139 & 10.937 &  9.561 &  8.671 &  7.768\\ 
0.55 & 3.59551 & -1.28094 & 0.500 & 1.254e+02 & 6.612e+01 & 3.030e+50 & 1.117e+53 & 9.931e+52 & 2.114e+53 & 11.582 & 10.343 &  8.958 &  8.086 &  7.270 & 6.406 & 5.769 & 5.617 & 11.577 & 10.370 &  9.039 &  8.188 &  7.376\\ 
0.60 & 3.61129 & -1.13877 & 0.500 & 1.068e+02 & 5.729e+01 & 6.108e+49 & 1.517e+53 & 1.044e+53 & 2.562e+53 & 11.029 &  9.789 &  8.455 &  7.638 &  6.905 & 6.080 & 5.443 & 5.312 & 10.990 &  9.790 &  8.514 &  7.711 &  6.984\\ 
0.65 & 3.62846 & -0.99758 & 0.500 & 9.305e+01 & 5.012e+01 & 0.000e+00 & 1.969e+53 & 1.070e+53 & 3.038e+53 & 10.439 &  9.230 &  7.959 &  7.217 &  6.580 & 5.736 & 5.113 & 5.002 & 10.374 &  9.204 &  7.995 &  7.245 &  6.597\\ 
0.70 & 3.64654 & -0.85948 & 0.500 & 8.117e+01 & 4.400e+01 & 0.000e+00 & 2.454e+53 & 1.082e+53 & 3.536e+53 &  9.832 &  8.696 &  7.499 &  6.819 &  6.253 & 5.415 & 4.810 & 4.708 &  9.726 &  8.651 &  7.511 &  6.821 &  6.233\\ 
0.75 & 3.66575 & -0.72461 & 0.500 & 7.089e+01 & 3.887e+01 & 0.000e+00 & 2.954e+53 & 1.090e+53 & 4.044e+53 &  9.124 &  8.120 &  7.022 &  6.420 &  5.926 & 5.123 & 4.555 & 4.459 &  9.054 &  8.117 &  7.052 &  6.422 &  5.886\\ 
0.80 & 3.68619 & -0.59368 & 0.500 & 6.254e+01 & 3.458e+01 & 0.000e+00 & 3.460e+53 & 1.093e+53 & 4.553e+53 &  8.349 &  7.543 &  6.558 &  6.032 &  5.594 & 4.866 & 4.346 & 4.258 &  8.345 &  7.585 &  6.603 &  6.039 &  5.553\\ 
0.85 & 3.70568 & -0.46899 & 0.500 & 5.501e+01 & 3.079e+01 & 0.000e+00 & 3.994e+53 & 1.081e+53 & 5.074e+53 &  7.634 &  7.040 &  6.157 &  5.681 &  5.270 & 4.632 & 4.153 & 4.074 &  7.686 &  7.091 &  6.190 &  5.684 &  5.243\\ 
0.90 & 3.72368 & -0.35030 & 0.500 & 4.790e+01 & 2.729e+01 & 0.000e+00 & 4.572e+53 & 1.036e+53 & 5.608e+53 &  7.015 &  6.605 &  5.809 &  5.367 &  4.974 & 4.404 & 3.959 & 3.886 &  7.089 &  6.644 &  5.822 &  5.366 &  4.958\\ 
0.95 & 3.73984 & -0.23755 & 0.500 & 4.124e+01 & 2.400e+01 & 0.000e+00 & 5.191e+53 & 9.679e+52 & 6.159e+53 &  6.475 &  6.204 &  5.480 &  5.070 &  4.695 & 4.191 & 3.790 & 3.723 &  6.536 &  6.226 &  5.477 &  5.065 &  4.689\\ 
1.00 & 3.75439 & -0.12957 & 0.500 & 3.487e+01 & 2.080e+01 & 0.000e+00 & 5.852e+53 & 8.648e+52 & 6.717e+53 &  5.992 &  5.835 &  5.173 &  4.797 &  4.446 & 3.987 & 3.625 & 3.568 &  6.081 &  5.859 &  5.175 &  4.795 &  4.444\\ 
1.05 & 3.76739 & -0.02597 & 0.500 & 2.864e+01 & 1.757e+01 & 0.000e+00 & 6.570e+53 & 7.092e+52 & 7.279e+53 &  5.583 &  5.499 &  4.888 &  4.541 &  4.213 & 3.779 & 3.446 & 3.394 &  5.659 &  5.515 &  4.888 &  4.538 &  4.209\\ 
1.10 & 3.77903 &  0.07324 & 0.500 & 2.193e+01 & 1.419e+01 & 4.481e+50 & 7.310e+53 & 5.254e+52 & 7.840e+53 &  5.226 &  5.190 &  4.620 &  4.298 &  3.992 & 3.573 & 3.263 & 3.212 &  5.264 &  5.186 &  4.615 &  4.291 &  3.983\\ 
1.15 & 3.78948 &  0.16746 & 0.500 & 1.467e+01 & 1.069e+01 & 1.960e+51 & 8.049e+53 & 3.331e+52 & 8.401e+53 &  4.911 &  4.898 &  4.363 &  4.061 &  3.774 & 3.392 & 3.107 & 3.060 &  4.951 &  4.901 &  4.369 &  4.066 &  3.774\\ 
1.20 & 3.79899 &  0.25725 & 0.500 & 8.141e+00 & 7.440e+00 & 3.944e+51 & 8.770e+53 & 1.525e+52 & 8.962e+53 &  4.617 &  4.623 &  4.119 &  3.837 &  3.566 & 3.218 & 2.959 & 2.916 &  4.652 &  4.627 &  4.134 &  3.849 &  3.573\\ 
1.25 & 3.80807 &  0.34281 & 0.500 & 2.394e+00 & 0.000e+00 & 5.794e+51 & 9.448e+53 & 2.806e+51 & 9.534e+53 &  4.344 &  4.363 &  3.890 &  3.625 &  3.371 & 3.056 & 2.822 & 2.782 &  4.370 &  4.368 &  3.911 &  3.644 &  3.383\\ 
1.30 & 3.81760 &  0.42605 & 0.500 & 0.000e+00 & 0.000e+00 & 6.669e+51 & 1.004e+54 & 0.000e+00 & 1.010e+54 &  4.084 &  4.110 &  3.667 &  3.420 &  3.182 & 2.902 & 2.694 & 2.657 &  4.110 &  4.115 &  3.693 &  3.445 &  3.200\\ 
1.35 & 3.82645 &  0.50530 & 0.500 & 0.000e+00 & 0.000e+00 & 7.989e+51 & 1.065e+54 & 0.000e+00 & 1.073e+54 &  3.848 &  3.873 &  3.457 &  3.226 &  3.001 & 2.754 & 2.570 & 2.535 &  3.884 &  3.884 &  3.490 &  3.257 &  3.027\\ 
1.40 & 3.83525 &  0.58081 & 0.500 & 0.000e+00 & 0.000e+00 & 9.578e+51 & 1.132e+54 & 0.000e+00 & 1.141e+54 &  3.623 &  3.648 &  3.260 &  3.048 &  2.838 & 2.613 & 2.448 & 2.415 &  3.668 &  3.659 &  3.294 &  3.077 &  2.862\\ 
1.45 & 3.84458 &  0.65226 & 0.500 & 0.000e+00 & 0.000e+00 & 1.106e+52 & 1.206e+54 & 0.000e+00 & 1.218e+54 &  3.404 &  3.432 &  3.076 &  2.885 &  2.693 & 2.483 & 2.337 & 2.306 &  3.461 &  3.446 &  3.109 &  2.910 &  2.711\\ 
1.50 & 3.85476 &  0.72048 & 0.500 & 0.000e+00 & 0.000e+00 & 1.331e+52 & 1.288e+54 & 0.000e+00 & 1.302e+54 &  3.227 &  3.227 &  2.896 &  2.721 &  2.539 & 2.366 & 2.232 & 2.201 &  3.278 &  3.242 &  2.934 &  2.752 &  2.569\\ 
\enddata
\end{deluxetable}
\end{landscape}
\end{document}